\documentclass[a4paper,12pt]{article}
\pdfoutput=1 

\usepackage{jheppub} 
\usepackage[export]{adjustbox}
\usepackage[T1]{fontenc} 
\usepackage{slashed}
\usepackage{pdflscape}
\usepackage{mathtools}
\usepackage{amsmath}
\usepackage{amsfonts}
\usepackage{pifont}
\usepackage{slashed}
\usepackage{cancel}
\usepackage{amssymb}
\usepackage{bbold}
\usepackage[dvipsnames]{xcolor}
\usepackage{multirow}
\usepackage[normalem]{ulem}
\usepackage{mathrsfs}

\usepackage{tikz}
\usepackage{tkz-euclide}
\usetkzobj{all}
\usetikzlibrary{decorations.pathmorphing}	
\tikzset{
    v/.style={decorate, decoration={snake, segment length=3mm, amplitude=0.75mm}, draw},
    f/.style={draw=black, postaction={decorate},
        decoration={markings,mark=at position .6 with {\arrow[very thick]{latex}}}},
    fb/.style={draw=black, postaction={decorate},
        decoration={markings,mark=at position .4 with {\arrowreversed[very thick]{latex}}}},
    fnar/.style={draw=black},
    g/.style={decorate, draw=black,
        decoration={coil,amplitude=3pt, segment length=3.5pt}},
    s/.style={dashed,draw=black, postaction={decorate},
        decoration={markings,mark=at position .55 with {\arrow[very thick]{latex}}}},
    sb/.style={dashed,draw=black, postaction={decorate},
        decoration={markings,mark=at position .55 with {\arrowreversed[draw=black,very thick]{latex}}}},
    snar/.style={dashed,draw=black,line width =1.25pt},
}

\newcommand{\al}[1]{\begin{align}\begin{aligned} #1 \end{aligned}\end{align}}
\newcommand{\be}{\begin{equation}}
\newcommand{\ee}{\end{equation}}
\newcommand{\bes}{\begin{equation*}}
\newcommand{\ees}{\end{equation*}}
\newcommand{\eL}{\epsilon_L}

\usepackage{amssymb}
\usepackage{pifont}
\newcommand{\cmark}{{ \ding{51}}}
\newcommand{\xmark}{{\ding{55}}}

\def\be{\begin{equation}}
\def\ee{\end{equation}}
\newcommand{\bea}{\begin{eqnarray}}
\newcommand{\eea}{\end{eqnarray}}

\title{ Minimal flavor violation in the see-saw portal }

\author[a]{Daniele Barducci}
\author[b]{Enrico Bertuzzo}
\author[c]{Andrea Caputo}
\author[c]{Pilar Hernandez}

\affiliation[a]{Dipartimento di Fisica Universit\`a degli Studi di Roma La Sapienza and INFN Sezione di Roma, Piazzale Aldo Moro 5, 00185, Roma, Italy}
\affiliation[b]{Instituto de Fisica, Universidade de Sao Paulo, C.P. 66.318, 05315-970 Sao Paulo, Brazil}
\affiliation[c]{Institut de F\'{i}sica Corpuscular - CSIC/Universitat de Val\`encia, Parc Cient\'ific de Paterna}

\emailAdd{daniele.barducci@roma1.infn.it}
\emailAdd{bertuzzo@if.usp.br}
\emailAdd{andrea.caputo@uv.es}
\emailAdd{pilar.hernandez@ific.uv.es}

\abstract{
We consider an extension of the Standard Model with two singlet leptons, with masses in the electroweak range, that induce neutrino masses via the see-saw mechanism, plus a generic new physics sector at a higher scale, $\Lambda$. We apply the minimal flavor violation (MFV) principle to the corresponding  Effective Field Theory ($\nu$SMEFT) valid at energy scales $E \ll \Lambda$. We  identify the irreducible sources of lepton flavor and lepton number violation at the renormalizable level, and apply the 
MFV ans\"atz to derive the scaling of the Wilson coefficients of the $\nu$SMEFT operators up to dimension six. We highlight the most important phenomenological consequences of this hypothesis in the rates for exotic Higgs decays, the decay length of the heavy neutrinos, and their production modes at present and future colliders. We also comment on possible astrophysical implications.}

\begin{document}

\maketitle

\section{Introduction}
\label{sec:intro}

The observed pattern of neutrino masses and oscillations parameters~\cite{Tanabashi:2018oca} calls for the existence of new physics (NP) beyond the Standard Model (SM). One of the simplest solution is to extend the SM with the right-handed (RH) chiral counterparts of the left-handed SM neutrinos, with which the new states can have Yukawa type interactions at the renormalizable level. Being electroweak (EW) singlets, the RH neutrinos $N_R$ (also dubbed sterile neutrinos) can have Majorana masses and  provide a mechanism that explains the lightness of the observed neutrinos in terms of a large hierarchy between the EW scale $v$ and the Majorana mass scale.
This is the essence of the see-saw mechanism~\cite{Minkowski:1977sc,Mohapatra:1979ia,Yanagida:1979as,GellMann:1980vs} which is parametrically expressed by the well-known relation
\be\label{eq:naive_see-saw}
m_{\nu_L} \propto \frac{y^2\, v^2}{M_{N_R}} \ ,
\ee
where $y$ and $M_{N_R}$ are the Yukawa coupling and the Majorana mass term for the RH neutrinos respectively.
In its original realization the mechanism assumes $M_{N_R}$ at around the Grand Unification Scale while the Yukawa coupling $y$ is an ${\cal O}(1)$ parameter. Low scale see-saw models, with RH neutrino masses at the EW scale, have recently received more attention. They can in fact explain the matter-antimatter asymmetry of the Universe via neutrino oscillations~\cite{Akhmedov:1998qx,Asaka:2005pn}, without introducing a severe fine tuning of the Higgs mass~\cite{Vissani:1997ys}. More interestingly they can also be tested for in beam dump experiments and at  colliders, see {\emph{e.g.}} \cite{Ferrari:2000sp,Graesser:2007pc,delAguila:2008cj,BhupalDev:2012zg,Helo:2013esa,Blondel:2014bra,Abada:2014cca,Cui:2014twa,Antusch:2015mia,Gago:2015vma,Antusch:2016vyf,Caputo:2016ojx,Caputo:2017pit}, possibly giving rise to spectacular signals such as displaced vertices. 

The presence of additional NP states at a scale $\Lambda \gg v, M_{N_R}$ can modify the phenomenological predictions of the see-saw model.  At low energy these effects  can be 
generically parametrized by an effective field theory (EFT) that contains a tower of higher dimensional operators ${\cal O}^{d}/ \Lambda^{d-4}$ with dimension $d>4$, that can induce new production and decay modes for the RH neutrinos, as well as new exotic Higgs decays~\cite{Graesser:2007yj,Graesser:2007pc,delAguila:2008ir,Aparici:2009fh,Caputo:2017pit,Yue:2017mmi,Butterworth:2019iff}. Sizable effects clearly arise only if $\Lambda$ is not too much higher than the EW scale. However, as it is well known, higher dimensional operators with a generic flavor structure, and suppressed by a scale $\Lambda \sim {\cal O}(1-10)\,$TeV, are grossly excluded by a variety of searches for flavor changing neutral currents (FCNC)  and lepton number violating (LNV) decays. For example the dimension six operator $(\bar L \sigma_{\mu\nu} e_R) H B^{\mu\nu}/ \Lambda^2$ induces at tree level the transition $\mu \to e \gamma$ for which the constraint from the MEG experiment~\cite{TheMEG:2016wtm} sets $\Lambda \gtrsim 6\times 10^4\,$TeV~\cite{Pruna:2014asa}.

The Minimal Flavor Violation (MFV) paradigm \cite{Chivukula:1987py,DAmbrosio:2002vsn} provides a  suppression for these processes derived from a symmetry principle. Briefly, it states that all flavor and charge-parity violating interactions in the EFT should be linked to the ones of the renormalizable Lagrangian. In practice, for the case of the quark sector of the SM this mechanism is implemented by promoting the Yukawa matrices to spurion fields with well-defined transformation properties under the flavor group in such a way that the full Lagrangian, including the non-renormalizable interactions, has the same global symmetry as the kinetic term~\cite{DAmbrosio:2002vsn}.
In the lepton sector however the still unknown mechanism that gives mass to the light neutrinos adds model-dependent spurions. For example, in the minimal see-saw model considered in this work, the leptonic spurions include the neutrino Yukawa coupling and the Majorana mass matrix for the RH neutrinos, $M_{N_R}$, which generally also acts as a source of lepton number breaking.  
Leptonic MFV in the context of the SMEFT has been first analyzed in~\cite{Cirigliano:2005ck,Davidson:2006bd,Branco:2006hz,Gavela:2009cd,Alonso:2011jd,Dinh:2017smk}, where the authors have identified the conditions under which one can expect measurable rates for LFV low-energy processes induced by higher dimensional operators. The main conclusion is that one needs a large separation between the scale of lepton number violation (LNV), for example $M_{N_R}$, and the scale of the higher dimensional operators that induce LFV processes, \textit{i.e.} $M_{N_R}\gg \Lambda$.  In this work we are instead interested  in  RH neutrinos at the EW scale, {\emph{i.e.}} $M_{N_R} \ll \Lambda $. The higher dimensional operators must therefore be built including also the RH neutrino fields, and it is precisely their phenomenology that we want to understand. In this sense our approach is then complementary to the one of~\cite{Cirigliano:2005ck,Gavela:2009cd}. 

The SMEFT extended to include the RH neutrino fields, that we will refer to as $\nu$SMEFT,  has been constructed up to $d=7$  in~\cite{Graesser:2007yj,Graesser:2007pc,delAguila:2008ir,Aparici:2009fh,Liao:2016qyd}. The smallness of neutrino masses are not compatible with large LFV effects from higher dimensional operators, unless the couplings of the higher dimensional operators are strongly hierarchical. For the $d=5$ operators, such a hierarchy has been shown to arise with the imposition of the MFV ans\"atz \cite{Graesser:2007yj}, as well as in the presence of an approximate $U(1)_L$ lepton number symmetry \cite{Caputo:2016ojx}.

In this paper we  systematically study the implications of the MFV ans\"atz on the scaling of the Wilson coefficients of all $d=5$ and $d=6$  operators involving RH neutrinos and SM fields, including quark bilinears. Particularly interesting for phenomenology is the scenario where the  textures of the neutrino spurions imply 
strong deviations from  the naive see-saw scaling of Eq.~\eqref{eq:naive_see-saw} \cite{Kersten:2007vk,Gavela:2009cd}, allowing for observable LFV effects compatible with the measured values of the light neutrino masses. We discuss in this context the implications  of MFV for the phenomenology of the RH neutrino states at present and future colliders. In particular, we study when prompt or displaced signatures can be expected from their decay, a property which is essential for experimental search strategies. We also qualitatively discuss the sensitivity of present and future experiments to the new physics scale $\Lambda$ via RH neutrino searches, stressing  the impact of the MFV ans\"atz.  Finally, we also briefly comment on astrophysical constraints, which are relevant when RH neutrinos masses lie in the keV to MeV range. 

The paper is organized as follows. In Sec.~\ref{sec:setting_stage} we fix our notation and discuss the global symmetries of the SM extended with an arbitrary number of RH neutrinos. In Sec.~\ref{sec:spurions} we review the  MFV ans\"atz and parametrize the various source of flavor breaking by ''composed spurion'' fields, while in Sec.~\ref{sec:mnu} we establish a connection between the masses of the active neutrinos and the higher dimensional operators that modify their mass spectrum. Sec.~\ref{sec:hierarchies_5} and Sec.~\ref{sec:hierarchies_6} discuss the scaling of the Wilson coefficients of the $d=5$ and $d=6$ operators under the MFV paradigm, while in Sec.~\ref{sec:pheno} we briefly highlight the more relevant phenomenological consequences. We then conclude in Sec.~\ref{sec:conclusions}.

\section{Setting the stage}
\label{sec:setting_stage}

We will work with the $\nu$SMEFT which is described by the following Lagrangian 

\al{\label{eq:EFT}
\mathcal{L}_{\rm \nu SMEFT} 
& \simeq  \mathcal{L}_{kin} - \bar{Q} Y_d H d_R- \bar{Q} Y_u \tilde{H} u_R- \bar{L} Y_e H e_R - \bar{L} Y_\nu \tilde{H} N_R - \frac{1}{2} \bar{N}_R^c M_N N_R + h.c.  \\ & + \frac{1}{\Lambda} \mathcal{L}_5 + \frac{1}{\Lambda^2} \mathcal{L}_6 + \dots
}
where $N_R^c=C \bar N_R^T$, $C=i\gamma^0 \gamma^2$, $\tilde H=i \sigma_2 H^*$. 
In Eq.~\eqref{eq:EFT} the terms in the first line describe the SM Lagrangian extended with renormalizable operators involving RH singlet fermions, while the terms $\mathcal{L}_{d>4}$ contain all the possible higher dimensional operators built out with the SM field content plus the RH neutrinos. In our analysis we will work up to dimension six, for which a complete list of operators can be found in~\cite{Grzadkowski:2010es, Graesser:2007yj,Graesser:2007pc,delAguila:2008ir,Aparici:2009fh,Liao:2016qyd}~\footnote{Ref.~\cite{Liao:2016qyd} provides also a list of dimension seven operators involving RH neutrino fields. The first list of $d=6$ operators including RH fields appeared in Ref. \cite{delAguila:2008ir}, but as pointed out in \cite{Liao:2016qyd} some of these were redundant.}.

Once the Yukawa interactions and the Majorana mass are switched off, the renormalizable part of the Lagrangian of  Eq.~\eqref{eq:EFT} has a global symmetry
\al{
 {\cal G}&= 
U(3)_{L} \times 
U(3)_{e} \times 
U(\mathcal{N})_{N} \times 
U(3)_{q} \times 
U(3)_{u} \times 
U(3)_{d} = \\
& = SU(3)^5 \times  U(1)^5 \times  SU(\mathcal{N})_{N} \times  U(1)_{N}  
}
where $\mathcal{N}$ is the number of RH neutrinos. We can rearrange the six $U(1)$ factors in different ways. One possible choice is to define as usual three factors to be the (global) hypercharge, baryon and lepton number. The three remaining factors can be chosen to be a Peccei-Quinn (PQ) like symmetry acting on $d_R$ and $e_R$, a phase acting on $e_R$ only (see {\emph{e.g.}}~\cite{DAmbrosio:2002vsn}) and an extra phase acting on the RH neutrinos. We choose to assign the same lepton number to all the RH neutrinos. There is however freedom in this choice. For example in the case ${\mathcal{N}}=2$ an interesting possibility would be to assign opposite charges to the two RH neutrinos as in the inverse see-saw model. We leave the discussion of this possibility for future work. Under these assumptions, let us analyze the group factors in more detail, focusing on the various sources of the breaking of the $U(1)$ symmetries.

\paragraph{Yukawa terms} Baryon and lepton number, together with the global version of the hypercharge, are respected by the Yukawa terms $Y_{u,d,e,\nu}$. The PQ symmetry $U(1)_{\rm PQ}$ is broken by $Y_d$ and $Y_e$, while $U(1)_{e}$ is broken by $Y_e$. Notice that the PQ symmetry plays an important role in flavor dynamics models with more than one Higgs doublet, since in that case it is possibile to assign a PQ charge to one of the two Higgs doublets, making then the Yukawa terms invariant under this symmetry~\cite{DAmbrosio:2002vsn}. Finally, $U(1)_{N}$ is broken by the neutrino Yukawa term.

\paragraph{Majorana mass} The Majorana mass term breaks both $U(1)_L$ and $U(1)_{N}$. 

\bigskip
With this said, we now focus on the flavor subgroup in the leptonic sector, 
\be
{\cal G}_L = SU(3)_{L} \times SU(3)_{e} \times SU(\mathcal{N})_{N} \times U(1)_\ell \times U(1)_{e} \times U(1)_{N} \ ,
\ee
and classify fields and spurions in terms of their transformations properties. From the field transformations
\be
L  \to e^{ i \alpha_\ell}\, V_L L, \qquad e_R  \to e^{i \alpha_\ell} \,V_e e_R, \qquad N_R  \to e^{i \alpha_\ell} \,V_N N_R,
\label{eq:field_transf}
\ee
where the $V_i$ matrices are unitary matrices belonging to $SU(3)_i$ and where we show only the lepton number transformation of parameter $\alpha_\ell$, the spurion transformations that leave the renormalizable part of the Lagrangian of Eq.~\eqref{eq:EFT} invariant read
\al{
Y_e \to V_L Y_e V_e^\dag, ~~~ Y_\nu \to  V_L Y_\nu V_N^\dag, ~~ M_N \to  e^{-2 i \alpha_\ell} \,V_N^* M_N V_N^\dag \ .
\label{eq:spurion_transf}
}

The Majorana mass matrix spurion $M_N$ transforms under $SU(\mathcal{N})_{N}$ as $\overline{\mathbf{S}}$, where $\mathbf{S}$ is the symmetric representation that can be constructed out of two fundamentals. For instance, $\overline{\mathbf{S}} = \mathbf{3}$ when $\mathcal{N} = 2$, or $\overline{\mathbf{S}} = \overline{\mathbf{6}}$ when $\mathcal{N} =3$. All together, the charge assignments under ${\cal G}_L$ are reported in Tab.~\ref{tab:charges_general}. A similar analysis can be performed for the quark sector. The analysis in this sector has been studied in detail and we refer the reader to Ref.~\cite{DAmbrosio:2002vsn} for a comprehensive discussion.

\begin{table}[tb]
\centering
\begin{tabular}{c|cccccc}
 & $SU(3)_{L}$ & $SU(3)_{e}$ & $SU(\mathcal{N})_N$ & $U(1)_\ell$ & $U(1)_{e}$  & $U(1)_{N}$\\
 \hline
 $L$ & $\mathbf{3}$ & $\mathbf{1}$ & $\mathbf{1}$ & $+1$ & $0$ & $0$\\
 $e_R$ & $\mathbf{1}$ & $\mathbf{3}$ & $\mathbf{1}$ & $+1$ & $+1$  & $0$ \\
 $N_R$ & $\mathbf{1}$ & $\mathbf{1}$ & $\mathbf{\mathcal{N}}$ & $+1$ & $0$ & $+1$ \\
 \hline\hline
 $M_N$ & $\mathbf{1}$ & $\mathbf{1}$ & $\overline{\mathbf{S}}$ & $-2$ & $0$  & $-2$\\
 $Y_e$ & $\mathbf{3}$ & $\overline{\mathbf{3}}$ & $\mathbf{1}$ & $0$ & $-1$  & $0$ \\
 $Y_\nu$ & $\mathbf{3}$ & $\mathbf{1}$ & $\overline{\mathbf{\mathcal{N}}}$ & $0$ & $0$  & $-1$ \\
 \hline
\end{tabular}
\caption{\label{tab:charges_general} Global charges of fields and spurions in the lepton sector.}
\end{table}

Without loss of generality we can now use the transformation of Eq.~\eqref{eq:field_transf} to go from Eq.~\eqref{eq:EFT} to a basis in which both $Y_e$ and $M_N$ are diagonal matrices with non negative entries.
In the same way we can also choose to go in a basis where $Y_d$ is diagonal with non negative entries and
$Y_u= V_{\rm CKM}^\dag m_u / v$, where $m_u$ is the diagonal matrix containing the physical up type quark masses and $V_{\rm CKM}$ is the Cabibbo-Kobayashi-Maskawa matrix with $u_L =  V_{\rm CKM} u_L^{\rm mass}$.

Note that we can decouple the sources of $SU({\cal N})_N$ and lepton number breaking by assuming that the Majorana mass matrix is proportional to the identity in flavor space as discussed in~\cite{Cirigliano:2005ck}. This reduces the flavor group from $SU(\mathcal{N})_N$ to $SO(\mathcal{N})_N$ thus making $V_N$ a real orthogonal matrix.

\section{Spurion parametrization}
\label{sec:spurions}

In the MFV paradigm the flavor structure of the non renormalizable operators  contained in ${\cal L}_{d>4}$ are to be built out of the irreducible sources of flavor breaking of the renormalizable Lagrangian in such a way that they are invariant under the full global symmetry group.
Applied to the quark sector this implies that higher dimensional operators should be built out with SM fields and the $Y_u$ and $Y_d$ spurion fields~\cite{DAmbrosio:2002vsn}.

The same paradigm applied to the lepton sector features a richer structure, due to the Majorana mass term that in general controls both the breaking of the flavor and of the lepton number symmetries, while the two Yukawa matrices, $Y_e$ and $Y_\nu$, act as a source of lepton flavor violation only~\cite{Cirigliano:2005ck}. We will impose the MFV hypothesis in the lepton sector by requiring that all the sources of lepton number and lepton flavor breaking of the $d>4$ operators are dictated by $M_N$, $Y_e$ and $Y_\nu$.
 To this end we now analyze in more detail the flavor breaking spurions reported in Tab.~\ref{tab:charges_general}. To consider only dimensionless quantities, we define the diagonal matrix
\be
\eL \equiv \frac{M_N}{\Lambda}\ ,
\label{eq:def_eps}
\ee
with the transformation properties of Eq.~\eqref{eq:spurion_transf}. 
In terms of $\eL$, the mass term of the RH neutrinos amounts to $ \Lambda \, \bar{N}_R^c \, \eL \, N_R$. In the $\eL \to 0$ limit we recover the $U(1)_\ell $ symmetry, {\emph{i.e.}} it is technically natural to take $\eL$ small. 
Note that the choice of Eq.~\eqref{eq:def_eps} connects with $\Lambda$ also the scale of lepton number breaking, which is due at the renormalizable level to the Majorana mass term $M_N$. Also, as already mentioned, when the sources of lepton flavor and lepton number breaking are decoupled, this spurion will be proportional to the identity matrix in flavor space. The transformation under $SU(\mathcal{N})_N$ becomes trivial, and $\eL$ is now a spurion controlling the breaking of lepton number only. 

In order to determine the scaling of operators with $d=5$ and $d=6$, it is convenient to define some objects with well defined transformation properties under the flavor groups built out combining the fundamental spurions of the quark and the lepton sector, $Y_{u,d,e,\nu}$ and $\eL$. The first useful class is made up by objects that transform as bifundamental under \textit{the same} $SU(3)$ flavor group. They are
\al{\label{eq:objects1}
\mathcal{S}_{LL^\dag} & \to V_L \, \mathcal{S}_{LL^\dag}\, V_L^\dag \ , & \qquad & & \mathcal{S}_{N N^\dag} & \to V_N \, \mathcal{S}_{N N^\dag}\, V_N^\dag \ ,\\
\mathcal{S}_{ee^\dag} & \to V_e\, \mathcal{S}_{ee^\dag}\, V_e^\dag \ , & \qquad & & \mathcal{S}_{qq^\dag}  &  \to V_q \,\mathcal{S}_{q q^\dag} \,V_q^\dag \ ,\\
\mathcal{S}_{uu^\dag}   & \to V_u \,  \mathcal{S}_{u u^\dag}\,  V_u^\dag \ , & \qquad & & \mathcal{S}_{dd^\dag}   & \to V_d \, \mathcal{S}_{d d^\dag} \, V_d^\dag\ .
}
We will also need objects transforming as bifundamental under \textit{different} $SU(3)$ flavor groups. They read
\al{\label{eq:objects2}
 \mathcal{S}_\nu & \to V_L \, \mathcal{S}_\nu \, V_N^\dag\ , & \qquad & & \mathcal{S}_{\nu^\dag} & \to V_N \, \mathcal{S}_{\nu^\dag}\,  V_L^\dag\ , \\
 \mathcal{S}_e & \to V_L\,  \mathcal{S}_e\,  V_e^\dag\ , & \qquad & & \mathcal{S}_{e^\dag} & \to V_e \, \mathcal{S}_{e^\dag} \, V_L^\dag \ ,\\
  \mathcal{S}_u & \to V_q\,  \mathcal{S}_u \, V_u^\dag\ , & \qquad & & \mathcal{S}_{u^\dag} & \to V_u\,  \mathcal{S}_{u^\dag} \, V_q^\dag \ ,\\
  \mathcal{S}_d & \to V_q\,  \mathcal{S}_d\,  V_d^\dag\ , & \qquad & & \mathcal{S}_{d^\dag} & \to V_d\,  \mathcal{S}_{d^\dag} \, V_q^\dag\ .
}
Finally, we introduce the objects that are responsible for the breaking of the lepton number symmetry. They transform as
\al{\label{eq:objects3}
 \mathcal{S}_{L^*L^\dag}  & \to e^{- 2 i \alpha_L }V_L^*\,  \mathcal{S}_{L^* L^\dag}\,  V_L^\dag\ , & \qquad & & \mathcal{S}_{N^* N^\dag} \to  e^{  - 2 i \alpha_L }V_N^*\,   \mathcal{S}_{N^* N^\dag} \, V_N^\dag\ .
}
We now want to write these objects in terms of the spurions in Tab.~\ref{tab:charges_general}. To this end, we define a general polynomial ${\cal F}_{\langle x,y\rangle}$ of two non commuting variables $x,y$ as
\be
 {\cal F}_{\langle x,y\rangle}  = \sum_{i=0}^{\infty} p_{i,\langle x,y\rangle}\ , 
 \label{eq:f_def}
\ee
where $p_{i,\langle x,y\rangle}$ indicates the sum of all possibile monomial factors, each with a generic complex coefficients, with total exponent $i$, taking into account that in general $[x,y]\ne0$. For example we have
\al{
p_{0,\langle x,y\rangle} & = a_{0} \\ 
p_{1,\langle x,y\rangle} & = a_{1_{(1)}} x + a_{1_{(2)}} y \\ 
p_{2,\langle x,y\rangle} & = a_{2_{(1)}} x^2 + a_{2_{(2)}} y^2 + a_{2_{(3)}} x y + a_{2_{(4)}}  y x \ .} 

The generalization to a polynomial of more than two variables is straightforward. In the case of a polynomial of one variable only, the expansion simply amounts to the usual $\mathcal{F}_{\langle x \rangle} = \sum a_n x^n$. The objects in Eq.~\eqref{eq:objects1} that transform as bifundamental under the same $SU(3)$ factor can thus be written in a compact way as
 \al{\label{eq:obj1}
\mathcal{S}_{LL^\dag}  &= \,  \mathcal{F}_{\langle Y_\nu Y_\nu^\dag, Y_e Y_e^\dag \rangle  } \ , & \qquad  & &  \mathcal{S}_{qq^\dag}  &= \,  \mathcal{F}_{\langle Y_u Y_u^\dag, Y_d Y_d^\dag \rangle  } \ ,  \\
 \mathcal{S}_{ee^\dag}  &= \mathcal{F}_{\langle Y_e^\dag \mathcal{G}_{\langle Y_e^\dag Y_e, Y_\nu Y_\nu^\dag \rangle  } Y_e  \rangle  }\ , & \qquad & &  \mathcal{S}_{dd^\dag}  &=  \mathcal{F}_{\langle Y_d^\dag \mathcal{G}_{\langle Y_d^\dag Y_d, Y_u Y_u^\dag \rangle  } Y_d  \rangle  }\ , \\
\mathcal{S}_{NN^\dag} & = \mathcal{F}_{\langle Y_\nu^\dag Y_\nu, \eL^* \eL, Y_\nu^\dag \mathcal{G}_{\langle Y_\nu Y_\nu^\dag, Y_e Y_e^\dag \rangle} Y_\nu \rangle} \ , & \qquad & &  \mathcal{S}_{uu^\dag}  &=  \mathcal{F}_{\langle Y_u^\dag Y_u, Y_u^\dag \mathcal{G}_{\langle Y_d Y_d^\dag, Y_u Y_u^\dag \rangle  } Y_u  \rangle  } \ , 
}
where ${\cal G}$ is defined in the same way as ${\cal F}$ of Eq.~\eqref{eq:f_def} with in general different coefficients.
Note that the expansion of all these terms starts with a term proportional to the identity in flavor space. 
Moving on to the objects in Eq.~\eqref{eq:objects2} that transform as bifundamental under different $SU(3)$ flavor groups, they can be written as
\al{\label{eq:obj2}
\mathcal{S}_{\nu} &=\mathcal{S}_{LL^\dag} Y_\nu\ , & \qquad & & \mathcal{S}_{\nu^\dag} &= Y_\nu^\dag \mathcal{S}_{LL^\dag}\ , \\
\mathcal{S}_{e} &= \mathcal{S}_{LL^\dag} Y_e \ , & \qquad & & \mathcal{S}_{e^\dag} &= Y_e^\dag \mathcal{S}_{LL^\dag} \ , \\
\mathcal{S}_{u} &= \mathcal{S}_{qq^\dag} Y_u \ , & \qquad & & \mathcal{S}_{u^\dag} &= Y_u^\dag \mathcal{S}_{qq^\dag}  \ , \\
\mathcal{S}_{d} &= \mathcal{S}_{qq^\dag} Y_d\ , & \qquad & & \mathcal{S}_{d^\dag} &= Y_d^\dag \mathcal{S}_{qq^\dag}  \ , \\
}
where now the expansion of each of the terms above starts with a term which is proportional to the respective Yukawa matrix. Here above we have used the definitions of Eq.~\eqref{eq:obj1} to keep track in a synthetic way of objects with defined transformation rules. In general the spurions that multiply the Yukawa matrices in Eq.~\eqref{eq:obj2} do not have the same expansion coefficients, $a_{i_{(j)}}$, as those in Eq.~\eqref{eq:obj1}. 

Finally, the expansion of the objects that explicitly break lepton number, Eq.~\eqref{eq:objects3}, reads
\al{
\mathcal{S}_{L^*L^\dag}  &=  \mathcal{S}_{LL^\dag}^*  \, Y_\nu^*\, \eL\, Y_\nu^\dag \, \mathcal{S}_{L L^\dag} \ , & \qquad & &\mathcal{S}_{N^* N^\dag} = \mathcal{S}^*_{N N^\dag} \eL \mathcal{S}_{NN^\dag} \ .
\label{eq:spurion_LN}
}

In what follows we will write everything in terms of these ''composed spurions'' and we will expand them at leading order in the $Y_\nu$ and $Y_e$ matrices. While we will have to find a connection with the observed values of neutrino masses and mixing parameters to determine the order of magnitude of the elements of $Y_\nu$, we can already determine the numerical size of the terms involving $Y_e$. Since we work in the basis in which $Y_e$ is diagonal with non negative entries, we have~\footnote{We work with $\langle H \rangle=174\,$GeV.}
\be\label{eq:numericalYe}
Y_e = \lambda_e^{\rm diag} \simeq \begin{pmatrix} 3 \times 10^{-6} & 0 & 0 \\ 0 &  6\times 10^{-4} & 0 \\ 0 & 0 &  10^{-2} \end{pmatrix}.
\ee
Analogously, in the down-quark sector we have
\be
Y_d = \lambda_d^{\rm diag} \simeq 
\begin{pmatrix}
3 \times 10^{-5} & 0 & 0 \\
0 & 6 \times 10^{-4} & 0 \\
0 & 0 & 2 \times 10^{-2}
\end{pmatrix}\ , 
\ee
while in the up-quark sector we obtain
\be\label{eq:Yu}
Y_u = V_{CKM}^\dag \lambda_u^{\rm diag} \simeq 
\begin{pmatrix}
 10^{-5} & - 2 \times 10^{-3} & 8 \times 10^{-3} \\
3 \times 10^{-6} & 7 \times 10^{-3} & 4 \times 10^{-2} \\
5 \times 10^{-8} & 3 \times 10^{-4} & 0.99
\end{pmatrix}\ .
\ee

An implicit assumption we are making in Eq.~\eqref{eq:obj1}, Eq.~\eqref{eq:obj2} and Eq.~\eqref{eq:spurion_LN} is that we can stop the polynomial expansion to some finite order in the spurion insertions. While this is clearly true for the spurions involving the charged lepton and quark Yukawa couplings, this requirement might not be satisfied  for the terms involving $Y_\nu$. One needs to check that bilinears constructed out of them like $Y_\nu Y_\nu^\dag$ have entries typically smaller than 1. This condition turns out to be satisfied for the range of RH neutrino masses we are interested in, and we will comment more on this in Sec.~\ref{sec:mnu}.

We now use the formal definition of the spurions to determine the scaling of the Wilson coefficients of the higher dimensional operators. They are summarized in Tab.~\ref{tab:dim5} for the $d=5$ operators and in Tab.~\ref{tab:D6_operators} for the $d=6$ operators.
For the $d=6$ case we only show the operators that contain one or more RH neutrino fields, while for $d=5$ we also show the Weinberg operator~\cite{Weinberg:1979sa}, due to its connection with the generation of neutrino masses. In Tab.~\ref{tab:D6_operators} the $\times$ symbol denotes the direct product between the two composite spurions. The flavor indices are contracted within the brackets. When more than one contraction of flavor indices is possible, we show only the less suppressed spurion combination.~\footnote{For instance, in the case of the ${\cal O}_{Ne}^6$ operator, we have an additional flavor combination in which the flavor index of $\bar{N}_R$ is contracted with the flavor index of $e_R$ via a ${\cal S}_\nu^\dag {\cal S}_e$ spurion (and the conjugate for the other flavor indices). This contribution is suppressed with respect to the one we show in Tab.~\ref{tab:D6_operators}.}
 In both tables we indicate with the subscripts $S$ and $A$ the  symmetric and antisymmetric flavor combinations. This comes from the fact that the operators ${\cal O}_{NH}^5$ and ${\cal O}_{NB}^5$ are symmetric and antisymmetric in the ${\cal N}$ flavor indices respectively. The same applies to the $O^6_{4N}$ operator. The latter also violates lepton number by four units and identically vanishes when all the four RH neutrinos are identical.
In Tab.~\ref{tab:dim5} and Tab.~\ref{tab:D6_operators} we also indicate whether the operators are expected to arise at tree level or at loop level in a generic ultraviolet (UV) completion as discussed in Refs.~\cite{Buchmuller:1985jz, Craig:2019wmo}. This will add an additional suppression factors $\propto (4 \pi)^{-2}$ to the corresponding spurion and it will be important when discussing the phenomenological implications of these operators. Note that we cannot write the $L$ and $B$ number violating operators in Tab.~\ref{tab:D6_operators}  in terms of the spurions introduced so far, since an additional source of $B$ number violation would be needed, see Sec.~\ref{sec:additional_spurions}.

 \begin{table}[tb]
  \begin{center}
 \begin{tabular}{c|c|c|c }
  & Operator & Scaling & Loop generated \\
 \hline
 ${\cal O}_{NH}^5$ & $ \bar{N}_R^c N_R H^\dag H$ & $[\mathcal{S}_{N^* N^\dag}]_S/2$ & \xmark \\
 ${\cal O}_{NB}^5$ & $\bar{N}_R^c \sigma^{\mu\nu} N_R B_{\mu\nu}$ & $[\mathcal{S}_{N^*N^{\dagger}}]_A$ & \cmark \\
 ${\cal O}_W^5$ & $(\bar{L}^c \epsilon H) (L \epsilon H)$ & $  [\mathcal{S}_{L^*L^\dag}]_S/2$ & \xmark \\
 \hline
 \end{tabular}
 \end{center}
 \caption{\label{tab:dim5} Dimension five operators constructed with the SM and the RH neutrino fields. We also show the scaling of their Wilson coefficients in terms of the spurions of Eq.~\eqref{eq:obj1} and Eq.~\eqref{eq:spurion_LN}, and whether they are generated at one loop in a general UV completion. The additional factor of $1/2$ is conventional and allows to simplify the mass matrix in Eq.~\eqref{eq:nu_mass_matrix}.} 
 \end{table}

\begin{table}[tb]
\begin{center}
\begin{tabular}{c|c|c|c}
\multicolumn{4}{c}{Operators involving the Higgs boson} \\
\hline \hline
  & Operator & Scaling & Loop generated \\
\hline
${\cal O}_{LNH}^6$ & $(\bar L \tilde H N_R)(H^\dag H)+h.c.$   & $\mathcal{S}_\nu$ & \xmark \\
${\cal O}_{LNB}^6$ & $(\bar L \sigma^{\mu\nu} N_R) B_{\mu\nu} \tilde H+h.c$& $\mathcal{S}_\nu$ & \cmark \\
${\cal O}_{LNW}^6$ &  $(\bar L \sigma^{\mu\nu} N_R) \sigma^a W_{\mu\nu}^a \tilde H+h.c$& $\mathcal{S}_\nu$ & \cmark \\
${\cal O}_{NH}^6$      &    $(\bar N_R \gamma^\mu N_R)(H^\dag i \overleftrightarrow{D}_\mu H)$  & $\mathcal{S}_{NN^\dag} $ & \xmark \\
${\cal O}_{NeH}^6$      &    $(\bar N_R \gamma^\mu e_R)(\tilde H^\dag i \overleftrightarrow{D}_\mu H)+h.c.$  & $\mathcal{S}_{\nu^\dag}  \mathcal{S}_e$ & \xmark \\
\multicolumn{4}{c}{}\\
\multicolumn{4}{c}{Operators unsuppressed by MFV} \\
\hline \hline 
  & Operator & Scaling & Loop generated \\
\hline
${\cal O}_{Ne}^6$      &    $(\bar N_R \gamma^\mu N_R)(\bar e_R \gamma_\mu e_R)$  & $\mathcal{S}_{NN^\dag} \times \mathcal{S}_{ee^\dag}$ & \xmark \\
${\cal O}_{Nu}^6$      &    $(\bar N_R \gamma^\mu N_R)(\bar u_R \gamma_\mu u_R)$  & $\mathcal{S}_{NN^\dag}  \times \mathcal{S}_{u u^\dag}$ & \xmark \\
${\cal O}_{Nd}^6$      &    $(\bar N_R \gamma^\mu N_R)(\bar d_R \gamma_\mu d_R)$  & $\mathcal{S}_{NN^\dag}  \times \mathcal{S}_{d d^\dag}$ & \xmark \\
${\cal O}_{Nq}^6$      &    $(\bar N_R \gamma^\mu N_R)(\bar q_L \gamma_\mu q_L)$  & $\mathcal{S}_{NN^\dag} \times \mathcal{S}_{q q^\dag}$ & \xmark \\
${\cal O}_{NL}^6$      &    $(\bar N_R \gamma^\mu N_R)(\bar L_L \gamma_\mu L_L)$  & $\mathcal{S}_{NN^\dag} \times \mathcal{S}_{LL^\dag}$ & \xmark \\
${\cal O}_{NN}^6$      &    $(\bar N_R \gamma^\mu N_R)(\bar N_R \gamma_\mu N_R)$  & $\mathcal{S}_{NN^\dag} \times \mathcal{S}_{NN^\dag} $ & \xmark \\
\multicolumn{4}{c}{}\\
\multicolumn{4}{c}{Other operators suppressed by MFV} \\
\hline \hline
  & Operator & Scaling & Loop generated \\
\hline
${\cal O}_{4N}^6$      &    $(\bar N^c_R N_R)(\bar N^c_R N_R)+h.c.$  & $[\mathcal{S}_{N^*N^\dag }\times \mathcal{S}_{N^* N^\dag}]_S$ & \xmark \\
${\cal O}_{Nedu}^6$      &    $(\bar N_R \gamma^\mu e_R)(\bar d_R \gamma_\mu u_R)$  & $\mathcal{S}_{\nu^\dag}\mathcal{S}_{e} \times \mathcal{S}_{d^\dag}\mathcal{S}_{u}$ & \xmark \\
${\cal O}_{NLqu}^6$ &  $(\bar N_R L) (\bar q_L u_R)+h.c$&  $\mathcal{S}_{\nu^\dag} \times \mathcal{S}_{u}$ & \xmark \\
${\cal O}_{LNqd}^6$ &  $(\bar L N_R) \varepsilon (\bar q_L d_R)+h.c$& $\mathcal{S}_\nu \times \mathcal{S}_{d}$ & \xmark \\
${\cal O}_{LdqN}^6$ &  $(\bar L d_R) \varepsilon (\bar q_L N_R)+h.c$& $\mathcal{S}_\nu \times \mathcal{S}_{d}$ & \xmark \\
${\cal O}_{LNLe}^6$ &  $(\bar L N_R) \varepsilon (\bar L e_R)+h.c$& $\mathcal{S}_\nu \times \mathcal{S}_e$ & \xmark \\
\multicolumn{4}{c}{}\\
\multicolumn{4}{c}{$L$ and $B$ violating four fermions operators} \\
\hline \hline
${\cal O}_{uddN}^6$      &    $(\bar u^c_R d_R \bar d^c_R) N_R+h.c.$ & \xmark &  \xmark\\
${\cal O}_{qqdN}^6$      &    $(\bar q^c_L \varepsilon q_L \bar d^c_R) N_R+h.c.$ & \xmark &  \xmark\\
\hline
\end{tabular}
\caption{Dimension six operators involving a RH neutrino $N_R$ \cite{Liao:2016qyd}. We also show the scaling of the Wilson coefficients in terms of the spurions of Eq.~\eqref{eq:obj1}, Eq.~\eqref{eq:obj2} and Eq.~\eqref{eq:spurion_LN}, and if they are generated at one loop in a general UV completion. The classification is useful in the discussion of the phenomenological implications of MFV, see Sections~\ref{sec:hierarchies_6} and~\ref{sec:pheno}.
}
\label{tab:D6_operators}
\end{center}
\end{table}

\section{Connection with the neutrino mass matrix}
\label{sec:mnu}

After electroweak symmetry breaking (EWSB) the operators ${\cal O}^5_{NH}$ and ${\cal O}^6_{LNH}$  contribute to the neutrino mass matrix. In this section we study these corrections assuming the MFV ans\"atz. For definitiveness, we work with $\mathcal{N} = 2$ RH neutrinos, {\emph{i.e.}} the minimal number of states with which is possible to generate the observed pattern of neutrino masses and mixings in the limit $\Lambda\rightarrow \infty$. By defining $n=(\nu_L, N_R^c)$ the mass Lagrangian ${\cal L}_{\rm mass}=-1/2\,\bar n^c \,{\cal M} \,n+h.c.$ can be written in terms of the following mass matrix
\be\label{eq:nu_mass_matrix}
\mathcal{M} = 
\left(
\begin{array}{ccc}
- [\mathcal{S}^*_{L^* L^\dag}]_S\, \frac{ v^2}{\Lambda}  & &Y_\nu v - \mathcal{S}_\nu \frac{v^3}{\Lambda^2} \\  
 & &\\
Y_\nu^T v - \mathcal{S}_\nu^T \frac{v^3}{\Lambda^2} & & \tilde{M}
\end{array}
\right)   \ ,
\ee
where we have defined
\be\label{eq:RH_nu_mass}
\tilde{M} = M_N  -   [\mathcal{S}_{N^* N^\dag}]_S \frac{v^2}{\Lambda} = \left(\eL -  [\mathcal{S}_{N^* N^\dag}]_S \frac{v^2}{\Lambda^2} \right) \Lambda\ .
\ee
The $\nu_L-\nu_L$ block in Eq.~\eqref{eq:nu_mass_matrix} is generated by the Weinberg operator ${\cal O}^5_W$. The $\nu_L -N_R$ block receives a $d=4$ contribution from the $\bar{L} \tilde{H} N_R$ operator, as well as a $d=6$ contribution from the operator ${\cal O}_{LHN}^6$. The RH neutrino mass matrix $\tilde M$ has a $d=4$ contribution, from $M_N$,  and a $d=5$ contribution from the operator ${\cal O}_{NH}^5$. The former dominates in the MFV ans\"atz, as can be easily derived from Eq.~\eqref{eq:def_eps} and~\eqref{eq:spurion_LN}
\begin{eqnarray}
M_N = \epsilon_L \Lambda \gg \epsilon_L \Lambda {v^2\over \Lambda^2} \propto[S_{N^* N^\dagger}]_S {v^2 \over \Lambda} \ .
\label{eq:pert_condition_M}
\end{eqnarray}

In order to compute the neutrino masses we diagonalize the matrix in Eq.~\eqref{eq:nu_mass_matrix} to first order in the active-sterile mixing, {\emph{i.e.}} assuming $Y_{\nu} v \ll M_N$ (a condition that, as we will see, will be always verified in the allowed region of parameter space). We get
\be\label{eq:nu_masses}
m_\nu \simeq  
 [\mathcal{S}^*_{L^* L^\dag}]_S\, \frac{ v^2}{\Lambda}  + v^2 \left(Y_\nu  - \mathcal{S}_\nu \frac{v^2}{\Lambda^2} \right) \tilde{M}^{-1} \left(Y_\nu^T - \mathcal{S}_\nu^T \frac{v^2}{\Lambda^2}\right)\ ,
\ee
where we redefined the phase of the LH neutrino fields to change the sign of the neutrino mass matrix $m_\nu$. Although not conventional, this choice allows to simplify the following equations.  With our assumption the matrix $\tilde{M}$ can be inverted perturbatively in powers of $v/\Lambda$. We obtain
\al{
\tilde M^{-1}\simeq 
\frac{1}{M_N} + \frac{1}{M_N}
[\mathcal{S}_{N^* N^\dag}]_S \frac{1}{M_N} \frac{v^2}{\Lambda} \ .
}
By considering again Eq.~\eqref{eq:spurion_LN} ({\emph{i.e.}}  $[\mathcal{S}_{N^* N^\dag}]_S = c\, \epsilon_L+\dots$), we can write this quantity as
\al{
\tilde M^{-1} = \frac{1}{M_N} \left(1+ 
c \frac{v^2}{\Lambda^2} \right)\ . 
}
Using this expression in Eq.~\eqref{eq:nu_masses} and taking $\mathcal{S}_\nu =  b Y_\nu$ as it follows from Eq.~\eqref{eq:spurion_LN}, we obtain an expression for the neutrino masses as an expansion in $v/\Lambda$:
\al{
m_\nu & \simeq  [\mathcal{S}^*_{L^* L^\dag}]_S\, \frac{ v^2}{\Lambda} + v^2 Y_\nu \frac{1}{M_N}Y_\nu^T \left(1+ (c-b) \frac{v^2}{\Lambda^2}\right)+ \dots \ .
\label{eq:mnu_pert}
}
We will now use the leading expression of the Weinberg operator computed according to Eq.~\eqref{eq:spurion_LN},
{\emph{i.e.}} $[\mathcal{S}^*_{L^* L^\dag}]_S = a Y_\nu \eL Y_\nu^T + \dots$, to write the neutrino mass matrix as
\be\label{eq:PMNS_neutrino}
m_\nu \simeq  v^2\, Y_\nu \frac{\left(1+(c-b) \frac{v^2}{\Lambda^2} \right)\mathbb{1}+ a\, \eL^2}{M_N}  Y_\nu^T = U^* m_{\nu}^{(d)} U^\dag \ .
\ee

In the last expression we have introduced the Pontecorvo-Maki-Nakagawa-Sakata (PMNS) matrix $U$~\cite{Pontecorvo:1957qd,Maki:1962mu}, and the matrix $m_\nu^{(d)}$ is diagonal with non negative entries. In the following, we will fix the phases
of the PMNS matrix to zero and the mixing angles to their latest fit~\cite{Tanabashi:2018oca}, unless otherwise specified.
Using Eq.~\eqref{eq:PMNS_neutrino} we can write
\be
Y_\nu \simeq \frac{1}{v} U^* \sqrt{\mu}\, \frac{\sqrt{M_N}}{\sqrt{\left(1+(c-b) \frac{v^2}{\Lambda^2} \right)\mathbb{1}+ a \eL^2}}\ ,
\ee
where $\sqrt{\mu}$ is a $3 \times 2$ matrix satisfying $\sqrt{\mu} \sqrt{\mu}^T = m_\nu^{(d)}$. This allows us to write a compact expressions for the various matrices involved. The most general form this matrix can take in the case of normal (NH) and inverted hierarchy (IH) is
\be
\sqrt{\mu_{{\rm NH}}} = 
\begin{pmatrix}
0 & 0 \\
-\sin z \sqrt{m_2}  & \pm \cos z  \sqrt{m_2} \\
  \cos z \sqrt{m_3}& \pm \sin z\sqrt{m_3} 
\end{pmatrix}\ ,\qquad 
\sqrt{\mu_{{\rm IH}}} = 
\begin{pmatrix}
-\sin z \sqrt{m_1}  & \pm \cos z  \sqrt{m_1} \\
 \cos z \sqrt{m_2}& \pm \sin z\sqrt{m_2} \\
 0 & 0 
\end{pmatrix}\ ,
\ee
where $m_i$ are the physical neutrino masses for the two hierarchies~\footnote{We remind that with two RH neutrinos in the NH case $m_{\nu_3}>m_{\nu_2}$ and $m_{\nu_1}=0$ while in the IH case $m_{\nu_2}>m_{\nu_1}$ and $m_{\nu_3}=0$. For the NH case we take $m_{\nu_2}=8.6\times 10^{-3}\,$eV and $m_{\nu_3}=4.9\times 10^{-2}\,$eV while for the IH we take $m_{\nu_1}=5.0\times 10^{-2}\,$eV and $m_{\nu_2}=5.1\times 10^{-2}\,$eV.}.
In the expressions above the angle $z$ can be taken complex. This is the so-called Casas-Ibarra parametrization~\cite{Casas:2001sr}, which can be written as
\be
\sqrt{\mu_{{\rm NH}}} = 
\begin{pmatrix}
0 & 0 \\
0 & \sqrt{m_2} \\
\sqrt{m_3} & 0
\end{pmatrix} \mathcal{R} \equiv \sqrt{m_{{\rm NH}}} \mathcal{R} \ ,
\qquad
\sqrt{\mu_{{\rm IH}}} = 
\begin{pmatrix}
0 & \sqrt{m_1} \\
 \sqrt{m_2}  & 0\\
0& 0
\end{pmatrix} \mathcal{R} \equiv \sqrt{m_{{\rm IH}}} \mathcal{R} \ ,
\label{Casas}
\ee
where $\mathcal{R}$ is a generic complex $2 \times 2$ matrix satisfying $\mathcal{R} \mathcal{R}^T = \mathbb{1}$
\be\label{eq:Rmatrix}
\mathcal{R} = 
\begin{pmatrix}
\cos z & \pm \sin z \\
- \sin z & \pm \cos z
\end{pmatrix}\ .
\ee
This form includes matrices with $\mathrm{det}\mathcal{R} = 1$ (proper rotations, to which the $+$ sign applies) and matrices with $\mathrm{det}\mathcal{R} = -1$ (to which the $-$ sign applies). A similar expression can be written in the inverted hierarchy case. Overall, for both hierarchies we write
\be
Y_\nu \simeq \frac{1}{v} U^* \sqrt{m}\, \mathcal{R} \frac{\sqrt{M_N}}{\sqrt{\left(1+(c-b) \frac{v^2}{\Lambda^2} \right)\mathbb{1}+a\eL^2}} \ ,
\label{eq:ynu}
\ee
where $\sqrt{m}$ is any of the two matrices defined in Eq.~\eqref{Casas}. The active sterile neutrino mixing thus reads

\al{\label{eq:mixing}
\theta_{\nu N}  & \simeq - v \left(Y_\nu - \mathcal{S}_\nu \frac{v^2}{\Lambda^2} \right)\, \frac{1}{\tilde{M}} + 2\, a \frac{v^2}{\Lambda} [\mathcal{S}_{L^* L^\dag}^*]_S \left(Y_\nu - \mathcal{S}_\nu \frac{v^2}{\Lambda^2} \right) v\, \frac{1}{\tilde{M}^2} \\
& \simeq -v\, Y_\nu \, \frac{1}{M_N} \left[ \left(1+(2 c - b) \frac{v^2}{\Lambda^2} \right) \mathbb{1} -2\, a\, \frac{v^2}{\Lambda^2} M_N^2 Y_\nu^T Y_\nu \frac{1}{M_N^2} + \dots \right] \\
& \simeq -U^* \sqrt{m} \mathcal{R} \frac{1}{\sqrt{M_N}}   + \dots \ .
}

In the limit of \textit{real} orthogonal $\mathcal{R}$ matrix it is easy to estimate the order of magnitude of the entries of $Y_\nu$. Taking $U$ and $\mathcal{R}$ with generic $\mathcal{O}(1)$ entries and degenerate masses for the RH neutrinos, $M_{N_1}=M_{N_2}=M_{N_{1,2}}$, we conclude that for both hierarchies the  entries of $Y_\nu$ scale as shown in Eq.~\eqref{eq:naive_see-saw},
\be\label{eq:Ynu_naive_estimate}
Y_\nu \sim \frac{\sqrt{M_{N_{1,2}} m_\nu}}{v} \sim 4 \times 10^{-8} \left(\frac{M_{N_{1,2}}}{1~\mathrm{GeV}} \right)^{1/2}\ .
\ee
For the numerical estimate we have assumed NH and $m_\nu = m_{\nu_3}$, but the expression is valid also for IH apart from small numerical factors. We have neglected corrections proportional to $(v/\Lambda)^2$ or $\eL^2$.
This naive estimate can be challenged by turning on the imaginary part of the $z$ angle of the $\mathcal{R}$ matrix. Writing $z = \alpha + i \gamma$ and taking the $\gamma \gg 1$ limit, we obtain
\be
\mathcal{R} \simeq \frac{e^{\gamma- i \alpha}}{2}
\begin{pmatrix}
1 & \pm i \\
- i & \pm 1
\end{pmatrix}\ .
\ee
We see that the imaginary part of the angle $z$ can break the naive see-saw scaling, and we thus need to modify Eq.~\eqref{eq:Ynu_naive_estimate}. The correct estimate in the $\gamma \gg 1$ limit is
\be\label{eq:Ynu_true_estimate}
Y_\nu \sim 2 \times 10^{-8} \, e^{\gamma-i \alpha} \, \left(\frac{M_{N_{1,2}}}{1~\mathrm{GeV}} \right)^{1/2}\ . 
\ee
The active-sterile mixing clearly has the same enhancement behavior and its entries read
\be\label{eq:increase}
 \theta_{i,\alpha}  \equiv \left(\theta_{\nu N} \right)_{i\alpha} \sim  7.2 \times 10^{-6} \, e^{\gamma-i \alpha} \, \left(\frac{1~\mathrm{GeV}}{M_{N_{1,2}}} \right)^{1/2}\ .
\ee
In the previous expression $\alpha = 1,\, 2$, $i = e, \mu, \tau$ and we show only the lowest order in $v/\Lambda$ and $\eL$. Higher orders can be easily taken into account, but for the range of masses we are interested in, and taking $\Lambda \gtrsim 1$ TeV, such corrections are at most of order $1 \% $ and we will neglect them. 

\begin{figure}[tb]
\begin{center}
\includegraphics[width=0.6\textwidth]{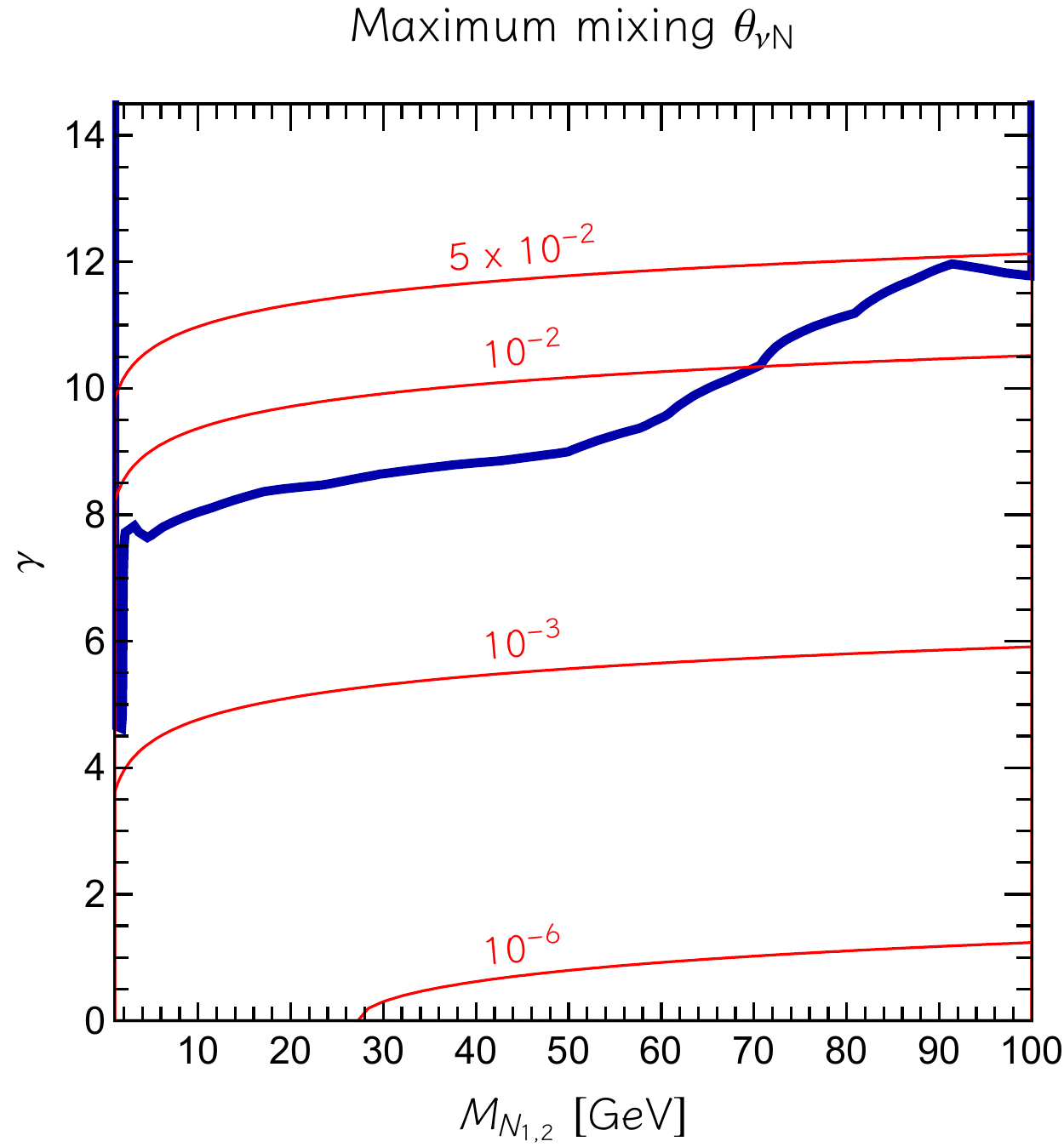}\hfill
\caption{\small Constraints on the mixing angles $\theta_{i}=\sum_{\alpha=1,2} |\theta_{i, \alpha}|^2$ in the $M_N-\gamma$ plane, where the masses of the RH neutrinos are taken degenerate with a value $M_{N_{1,2}}$. The region above the colored thick line is excluded by the bounds on active-sterile mixing angles~\cite{Liventsev:2013zz, Aaij:2016xmb, Abreu:1996pa}.  We show only the most restrictive bound for NH hierarchy, \textit{i.e.} the one coming from the $\theta_{\mu}$. The red thin lines show the corresponding maximum value of the active-sterile mixing angle computed according to Eq.~\eqref{eq:mixing}.}
\label{fig:active_sterile_bounds}
\end{center}
\end{figure}

The mixing angles are constrained by a variety of experimental searches and large value of $\gamma$ are ruled out. Using the bounds on $\theta_{i}=\sum_{\alpha=1,2} |\theta_{i, \alpha}|^2$, with $i=e,\mu,\tau$, reported in~\cite{Liventsev:2013zz,Aaij:2016xmb, Abreu:1996pa} we show in Fig.~\ref{fig:active_sterile_bounds} the allowed region in the $M_N-\gamma$ plane, assuming degenerate masses for $N_1$ and $N_2$ and neglecting the small $\Lambda$ dependence. For concreteness we show only the most stringent bound, coming from  $\theta_{\mu}$. The bound applies for both hierarchies~\footnote{Strictly speaking, in the case of IH the most stringent bound for masses below $70$ GeV is the one coming from $\theta_{e}$. Numerically however the bound is only slightly more stringent than the $\theta_{\mu}$ one. For simplicity we therefore only show  the latter.}. For $M_{N_{1,2}} = (1\div100)\,$GeV we see that values of $\gamma$ up to $\sim$ 8 are allowed by existing constraints. We also show the maximum value of the active-sterile mixing matrix computed according to Eq.~\eqref{eq:mixing}. As we can see, for low values $\gamma \sim 1\div 2$ the maximum mixing is of order $10^{-6}$, and it increases until a maximum value of $5 \%$ for $\gamma \simeq 12$ and $M_{N_{1,2}} \simeq (80 \div 100)$ GeV.
These values leave us safely within the range of the perturbative diagonalization performed to derive Eq.~\eqref{eq:nu_mass_matrix}. These values for the $\theta_{\nu N}$ matrix will be important in the spurion discussion in Sec.~\ref{sec:hierarchies_5} and Sec.~\ref{sec:hierarchies_6}.

With this information we can go back to a point already raised in Sec.~\ref{sec:spurions}. Implicit in the definitions of Eq.~\eqref{eq:objects1}, Eq.~\eqref{eq:objects2} and Eq.~\eqref{eq:objects3} is the fact that we can expand in $Y_\nu$, so schematically one requires $Y^{ij}_\nu \lesssim 1$. For this to be true we need 
\be
e^\gamma \lesssim 0.5 \times 10^8 \left( \frac{1~\mathrm{GeV}}{M_{N_{1,2}}} \right)^{1/2}\ .
\ee
This gives $\gamma \lesssim 15\,(17)$ for $M_{N_{1,2}}=1\;$(100) GeV. We conclude that whenever the experimental bounds on the active-sterile mixing are satisfied the expansion holds and we can keep only the lowest order terms in the spurion expansions.

\section{Hierarchies between the composed spurions: d=5 operators}
\label{sec:hierarchies_5}

We now use the parametrization of Sec.~\ref{sec:spurions} and the constraints of Sec.~\ref{sec:mnu} to express the composed spurions of Eq.~\eqref{eq:objects1}, Eq.~\eqref{eq:objects2} and Eq.~\eqref{eq:objects3} in terms of fundamental spurions. 
The aim is to understand their order of magnitude and the relative importance of the Wilson coefficients of the $d=5$ operators.


\subsection{Operator ${\cal O}^5_{NH}$}

The ${\cal O}^5_{NH}$ operator has a Wilson coefficient proportional to the symmetric part of the $\mathcal{S}_{N^* N^\dag}$ spurion. Working at the next-to-leading order in $Y_\nu$ or $\epsilon_L$, we have
\be
\mathcal{S}_{N^* N^\dag} = a_0\,  \eL + a_1 Y_\nu^T Y_\nu^* \eL + a_2 \eL Y_\nu^\dag Y_\nu +a_3 \eL \eL^* \eL + \dots,
\label{eq:d5_higgs_expansion}
\ee
where the coefficients $a_i$ are of order unity and, as already pointed out, the entries of the bilinears built from $Y_\nu$ are somewhat smaller than unity.
The relevant combination for the ${\cal O}_{NH}^5$ operator is the flavor symmetric, and we thus have
\al{\label{eq:spurion_ONH5}
[\mathcal{S}_{N^*N^\dag}]_S & \simeq \left( a_0 + a_3 \eL \eL^*\right) \eL + \frac{a_1+a_2}{2}  \left( Y_\nu^T Y_\nu^* \eL + \eL Y_\nu^\dag Y_\nu \right)  \\
		& \simeq a_0 \frac{M_N}{\Lambda}  + a_3 \frac{M_N^3}{\Lambda^3}+  \frac{a_1+a_2}{2 \, v^2 \, \Lambda}   \left(M_N^{1/2} \mathcal{R}^T \tilde{m}_\nu \mathcal{R}^* M_N^{3/2} + M_N^{3/2} \mathcal{R}^\dag \tilde{m}_\nu \mathcal{R} M_N^{1/2} \right) \ ,
}
where we have defined $\tilde{m}_\nu \equiv \sqrt{m}^T \sqrt{m}$. In the expression for $Y_\nu$ we kept only the leading terms in $v/\Lambda$ and $\eL$. For both normal and inverted hierarchy this matrix reads
\be
\label{eq:tilde_m}
\tilde{m}_\nu = 
\begin{pmatrix}
m_{heavy} & 0 \\
0 & m_{light}
\end{pmatrix}\ ,
\ee
where $m_{heavy}=m_{3} (m_2)$ and $m_{light}=m_{2} (m_1)$ in the normal (inverted) case. 
Using this expression in Eq.~\eqref{eq:spurion_ONH5} together with Eq.~\eqref{eq:Rmatrix} we obtain
\al{\label{eq:spurion_NstarNdag}
[\mathcal{S}_{N^*N^\dag}]_S & \simeq \frac{M_N}{\Lambda} \left[a_0 + \frac{a_1+a_2}{v^2} M_N \bigg( \bar{m}  \cosh(2\gamma) \mathbb{1} + \Delta m  \cos(2\alpha) \sigma_3 \bigg) + a_3 \frac{M_N^2}{\Lambda^2}\right]  \\
& \qquad{} \pm \frac{(a_1+a_2)\sqrt{M_1 M_2}}{v^2\, \Lambda} \bigg(\bar{M} \, \Delta m  \sin(2\alpha) - i \Delta M\,  \bar{m} \sinh(2\gamma) \bigg) \sigma_1\ ,
}
where $\sigma_1$ and $\sigma_3$ are the Pauli matrices. To simplify the equation we have defined $\bar{m} = (m_{light} + m_{heavy})/2$ and $\Delta m =( m_{heavy} - m_{light})/2$ for the light neutrinos, $\bar{M} = (M_1+M_2)/2$ and $\Delta M=(M_2-M_1)/2$ for the RH neutrinos. As expected, we see that the terms in Eq.~\eqref{eq:spurion_ONH5} arising from the neutrino Yukawa matrix are suppressed by the mass scale of the neutrinos $\bar m$ or their mass difference $\Delta m$. As a consequence, only the terms exponentially enhanced by the imaginary part of the $z$ rotation angle in the ${\cal R}$ matrix, see Eq.~\eqref{eq:Rmatrix}, can be important when $\gamma \gg 1$. 
By taking the limit of degenerate RH neutrino masses $\Delta M\to 0$ we obtain
\be
\label{eq:spurion_SNsNdag}
 [\mathcal{S}_{N^*N^\dag}]_S^{NH}  \simeq \frac{\bar{M}}{\Lambda}   \left( a_0 + a_3 \frac{\bar{M}^2}{\Lambda^2} +  10^{-13}\, (a_1+ a_2 )\cosh(2\gamma) \frac{\bar{M}}{100\,{\rm GeV}}\right) \times \mathbb{1} \ ,
 \ee
 where we have neglected entries suppressed by the neutrino mass scale or mass difference not proportional to an hyperbolic function of $\gamma$. The same expression is valid for both hierarchies apart from $\mathcal{O}(1)$ factors. In the case of non degenerate RH neutrino mass there will be an off-diagonal exponentially enhanced term proportional to $\sigma_1$. Note that the term proportional to the Yukawa coupling grows with $\bar{M}$ and $\gamma$. Considering the experimentally allowed region in Fig.~\ref{fig:active_sterile_bounds}, we see that for $\bar{M} \simeq 100~\mathrm{GeV}$ and $\gamma \simeq 12$ this term can be at most an order $10^{-3}$ correction to the leading contribution for both hierarchies. 
Importantly, and as already noticed in \cite{Graesser:2007yj}, the operator ${\cal O}_{NH}^5$ is not suppressed by a Yukawa at leading order,
although it turns out to be suppressed by $\eL$ also in the $M_N \sim \mathbb{1}$ limit in which the  $SU(2)_N$ factor in ${\cal G}_L$ reduces to $SO(2)_N$.

\subsection{Operator ${\cal O}^5_{NB}$}

Moving to the ${\cal O}_{NB}^5$ operator, the relevant contribution is now the antisymmetric one, which reads
\al{\label{eq:spurion_ONB5}
[\mathcal{S}_{N^*N^\dag}]_A & \simeq \frac{a_1 - a_2}{2} \left( Y_\nu^T Y_\nu^* \eL - \eL Y_\nu^\dag Y_\nu \right) \\
						& \simeq \frac{a_1 - a_2}{2\, v^2\, \Lambda} \left(M_N^{1/2} \mathcal{R}^T \tilde{m}_\nu \mathcal{R}^* M_N^{3/2}- M_N^{3/2} \mathcal{R}^\dag \tilde{m}_\nu \mathcal{R} M_N^{1/2} \right)\ .
}
Again, we keep only terms at the smallest order in $v/\Lambda$ and $\eL$. Using the expression for $\tilde{m}_\nu$ of Eq.~\eqref{eq:tilde_m} together with Eq.~\eqref{eq:Rmatrix} we obtain the simple expression
\be
[\mathcal{S}_{N^*N^\dag}]_A  \simeq \pm \frac{(a_1 - a_2)\sqrt{M_1 M_2}}{ v^2\, \Lambda} \left[\bar{m} \, \bar{M}\sinh(2\gamma) + i \Delta m \, \Delta M \sin(2\alpha) \right] \sigma_2\ ,
\ee
where $\sigma_2$ is the second Pauli matrix. 
Note that in the $SU(2)_N\to SO(2)$ limit the second term vanishes, and we are left with a dependence on $\gamma$ only. In this limit we obtain

\be\label{dipoleCon}
[\mathcal{S}_{N^* N^\dag}]_A \simeq \pm \,   10^{-13} (a_1-a_2)\, \sinh(2\gamma)\, \left( \frac{\bar{M}}{\Lambda}\right) \frac{\bar M}{100\,{\rm GeV}}  \times \sigma_2 \ ,
\ee
where the same expression is valid for both hierarchies, apart from $\mathcal{O}(1)$ numerical factors, and the high suppression due to the active neutrino mass scale is evident. We conclude that RH neutrinos production processes mediated by the ${\cal O}^5_{NB}$ operator, such as $pp \to \gamma, Z \to N_1 N_2$, turn out the be completely irrelevant if MFV is imposed, while heavy to light decay as {\emph{e.g.}}, $N_2 \to N_1 \gamma, N_1 Z$ will have a highly suppressed partial width.

\subsection{Operator ${\cal O}^5_{W}$}

The Weinberg operator carries the spurion $[\mathcal{S}_{L^* L^\dag}]_S$, which starts its expansion as
\begin{equation}
	[\mathcal{S}_{L^* L^\dag}]_S \simeq Y_\nu^*\, \eL\, Y_\nu^\dag  + \dots \simeq \frac{1}{v^2} U \sqrt{m} \mathcal{R}^* \frac{M_N^2}{\Lambda\, \sqrt{\mathbb{1}+a \eL^2}} \mathcal{R}^\dagger \sqrt{m} U^T \ .
\end{equation}
In the limit of degenerate RH neutrino masses the Casas-Ibarra matrix disappears from the expression. For non-degenerate RH neutrinos, however, there is a residual $\gamma$ dependence that could make this term large. The potentially large term can be easily isolated by writing $M_N = \bar{M} \mathbb{1} - \Delta M \sigma_3$. To leading order in $v/\Lambda$ and $\eL$ we obtain

\be
 [\mathcal{S}_{L^* L^\dag}]_S \simeq \frac{\bar{M}^2 + \Delta M^2}{v^2\, \Lambda } m_\nu - \frac{2 \Delta M\, \bar{M}}{v^2\, \Lambda} U^* \sqrt{m} \mathcal{R} \sigma_3 \mathcal{R}^T \sqrt{m}^T U^\dag\ .
\label{eq:ws}
\ee
Interestingly, the spurion of the Weinberg operator is not simply proportional to the light neutrino masses for non-degenerate heavy neutrinos.  This in a explicit demonstration that there can be lepton number breaking effects that the neutrino mass is not sensitive to at tree level, as well known in the so-called extended see-saw scenarios \cite{LopezPavon:2012zg}.  Indeed, the second term in Eq.~\eqref{eq:ws} enhanced by $\gamma$ gets cancelled against the second term in Eq.~(\ref{eq:mnu_pert}) in the neutrino mass matrix, but could be parametrically much larger than the latter if $\gamma$ and $\Delta M$ are large enough. On the contrary, for $\Delta M=0$  and  large $\gamma$ , one can show that there is effectively an approximate lepton number symmetry (that assigns opposite lepton number charges to the two $N_R$ fields), which suppresses the neutrino mass and any other lepton number breaking effect, but this not the case for non-degenerate neutrinos. 

\section{Hierarchies between the composed spurions: d=6 operators}
\label{sec:hierarchies_6}

We now analyze the scaling of the Wilson coefficients of the $d=6$ operators. Unlike what we did in Sec.~\ref{sec:hierarchies_5}, for these operators we find more convenient to organize the discussion in terms of the spurions. We  follow the classification outlined in Tab.~\ref{tab:D6_operators}. We start from the operators involving the Higgs field because they will be the most relevant for the phenomenological considerations of Sec.~\ref{sec:pheno}. Inspecting Tab.~\ref{tab:D6_operators} we immediately see that we can classify them in three categories: (i) operators that scale like $\mathcal{S}_\nu$,namely ${\cal O}_{LHN}^6$, ${\cal O}_{LNB}^6$ and ${\cal O}_{LNW}^6$, (ii) operators that scale like $\mathcal{S}_{NN^\dag}$, namely ${\cal O}_{NH}^6$, and (iii) operators that scale like $\mathcal{S}_{\nu^\dag} \mathcal{S}_e$, namely ${\cal O}_{NeH}^6$. We will then comment on the scaling of the other spurions.

\subsection{Spurion $\mathcal{S}_\nu$}
\label{sec:snu}

Let us start with the spurion $\mathcal{S}_\nu$. From Eq.~\eqref{eq:obj2}  we can write it as
\be\label{eq:Snu}
\mathcal{S}_\nu = a_0 Y_\nu + a_1 Y_\nu Y_\nu^\dag Y_\nu + a_2 Y_e Y_e^\dag Y_\nu + \dots
\ee
We will now show that $\mathcal{S}_\nu \simeq Y_\nu$ apart from corrections at most of order $\mathcal{O}(10^{-3})$. To see this it is easier to rewrite Eq.~\eqref{eq:Snu} in terms of the mixing angle matrix $\theta_{\nu N}$ of Eq.~\eqref{eq:mixing}. We obtain
\be
\mathcal{S}_\nu = \left[ a_0 + a_1\, \frac{\theta_{\nu N} M_N^2 \theta_{\nu N}^\dag}{v^2} + a_2\, \lambda_e^2 \right] Y_\nu + \dots \ .
\ee
Remembering now that in the experimentally allowed region of Fig.~\ref{fig:active_sterile_bounds} the entries of the $\theta_{\nu N}$ matrix are at most of order $5\%$, we conclude that we can write
\be
\mathcal{S}_\nu \lesssim  \left[ a_0 + a_1 \times 10^{-3} \left(\frac{M_{N_{1,2}}}{100\, \mathrm{GeV}}\right)^2 + a_2\times 10^{-4} \right] Y_\nu + \dots
\ee
as claimed above. In what follows we will always take $\mathcal{S}_\nu \simeq Y_\nu$. To obtain an approximate expression for $Y_\nu$ (and therefore for the spurion $\mathcal{S}_\nu$)  we write it in the limit of degenerate light and heavy neutrinos masses. The light neutrino mass scale will be denoted by $m$, and the RH neutrino mass scale with $M$. To simplify further the expressions, we will take the limit $\theta_{13} \to 0$ and $\theta_{12} \to \pi/4$ in the PMNS matrix. In the case of normal hierarchy we obtain
\be
Y_\nu^{({\rm NH})} \simeq \cosh(\gamma - i \alpha) \frac{\sqrt{m\, M}}{v} 
\begin{pmatrix}
- \frac{i}{\sqrt{2}} & \pm \,\frac{1}{\sqrt{2}} \\
- i\,  \tilde{c}_{23} & \pm\,  \tilde{c}_{23} \\ 
i \, \tilde{s}_{23} & \mp \tilde{s}_{23}
\end{pmatrix}\ ,
\ee
where we have defined $\tilde{c}_{23} \equiv c_{23}/ \sqrt{2} + i s_{23}$ and $\tilde{s}_{23} = s_{23}/\sqrt{2} - i c_{23}$ in terms of $s_{23} = \sin\theta_{23}$ and $c_{23} = \cos\theta_{23}$, with $\theta_{23}$ the atmospheric angle of the PMNS matrix. In the case of inverted hierarchy we instead obtain
\be
Y_\nu^{({\rm IH})} \simeq \cosh(\gamma - i \alpha) \frac{\sqrt{m\, M}}{v} \, e^{-i \pi/4} 
\begin{pmatrix}
1 & \pm\, i \\
i \, c_{23} & \mp\, c_{23}\\
-i\, s_{23} & \pm\, s_{23}
\end{pmatrix}\ .
\ee


\subsection{Spurion $\mathcal{S}_{NN^\dag}$}\label{sec:SNNdag}

We now analyze the spurion $\mathcal{S}_{NN^\dag}$, which appears in the operator ${\cal O}_{NH}^6$. At leading order this spurion can be expanded as
\al{
\mathcal{S}_{NN^\dag} & \simeq a_0\mathbb{1}+a_1 \epsilon^*_L \epsilon_L + a_2 Y_\nu^\dag Y_\nu + \dots \\
				& \simeq a_0 \mathbb{1} + a_1 \frac{M_N^2}{\Lambda^2} + \frac{a_2}{v^2} M_N^{1/2} \mathcal{R}^\dag \tilde{m}_\nu \mathcal{R} M_N^{1/2} + \dots \\
				& \simeq a_0 \mathbb{1} + a_1 \frac{M_N^2}{\Lambda^2} + \frac{M_N}{v^2} \bigg( \bar{m} \cosh(2\gamma) \mathbb{1} + \Delta m \cos(2\alpha) \sigma_3 \bigg) \\
				& \qquad {} \pm \frac{\sqrt{M_1 M_2}}{v^2} \bigg( \Delta m \sin(2\alpha) \sigma_1 - \bar{m} \sinh(2\gamma) \sigma_2 \bigg)\ \simeq \mathbb{1}\Big(a_0 + a_1 \frac{M_N^2}{\Lambda^2}\Big).
}
where we omit to write terms of order $O(10^{-15})\times \cosh(2\gamma)$ which are always negligible for any allowed value of the angle $\gamma$.

\subsection{Spurion $\mathcal{S}_{\nu^\dag} \mathcal{S}_e$}

We now move on to the spurion combination $\mathcal{S}_{\nu^\dag} \mathcal{S}_e$ appearing in the scaling of the operator ${\cal O}_{NeH}^6$. As shown in Sec.~\ref{sec:snu}, the dominant term in the expansion of the spurion $\mathcal{S}_\nu$ is given by $Y_\nu$. At the same time, it is clear from Eqs.~\eqref{eq:numericalYe} and~\eqref{eq:obj2} that the dominant term in the expansion of $\mathcal{S}_e$ is given by $Y_e$~\footnote{As for the case of the spurion $\mathcal{S}_{NN^\dag}$ in Sec.~\ref{sec:SNNdag}, the dominant term in the expansion of the spurion $\mathcal{S}_{LL^\dag}$ appearing in Eq.~\eqref{eq:obj2} is the one proportional to the identity.}.Putting all together we conclude that 
\be
	\mathcal{S}_{\nu^\dag}  \mathcal{S}_e \simeq a_0\, Y^{\dagger}_{\nu}\, Y_e + \dots
\ee
Given the small entries in the diagonal $\lambda_e$ matrix, Eq.~\eqref{eq:numericalYe},  we see that the entries of $ \mathcal{S}_{\nu^\dag}  \mathcal{S}_e$ are suppressed with respect to $Y_\nu$, Eq~\eqref{eq:ynu}. The minimal suppression, by a factor of order $10^{-2}$, involves the charged leptons of the third generation.

\subsection{Additional spurions}\label{sec:additional_spurions}

Using the results presented above it is immediate to compute the leading terms in the expansion of the Wilson coefficients of the remaining operators in Tab.~\ref{tab:D6_operators}. More specifically, the dominant term for all the unsuppressed operators is proportional to the identity
\be
\mathcal{S}_{X X^\dag } \simeq {\mathcal O}(1), \;\;\; X=e,u,d,q,L,N.
\ee
The remaining suppressed operators  in Tab.~\ref{tab:D6_operators} scale as
\al{
[\mathcal{S}_{N^*N^\dag }\times \mathcal{S}_{N^* N^\dag}]_S & \simeq \left(\frac{M_N}{\Lambda}\right) \times \left(\frac{M_N}{\Lambda}\right) \ , 
&  
\mathcal{S}_{\nu^\dag}\mathcal{S}_{e} \times \mathcal{S}_{d^\dag}\mathcal{S}_{u} & \simeq \left(Y_\nu^\dag Y_e\right) \times \left(Y_d^\dag Y_u \right) \ , \\
\mathcal{S}_{\nu^\dag} \times \mathcal{S}_{u} & \simeq \left(Y_\nu^\dag\right) \times \left(Y_u \right) \ , 
& 
\mathcal{S}_\nu \times \mathcal{S}_{d} & \simeq \left(Y_\nu \right) \times \left( Y_d\right)\ , 
}
and 
\be
\mathcal{S}_\nu \times \mathcal{S}_e \simeq \left(Y_\nu \right) \times \left(Y_e \right)\ .
\ee

Finally, the Wilson coefficient of the operators that violate both $B$ and $L$ number cannot be written solely in terms of the spurions we have introduced. Additional sources of baryon and lepton number violation are needed~\footnote{As an example, we can consider the operator ${\cal O}_{uddN}^6$. This can be obtained at tree level introducing a Yukawa term like
\be
\mathcal{L} = \lambda_\phi \bar d^c_R d_R \tilde{\phi} + \lambda_\phi^\prime \bar u^c_R N_R \tilde{\phi}^\dag\ ,
\ee
with $\tilde{\phi}$ a new scalar field with quantum numbers $(\mathbf{3},\mathbf{1})_{-2/3}$ under the SM gauge group. Integrating out $\tilde{\phi}$ at tree level one produces the operator ${\cal O}_{uddN}^6$, with the Yukawa couplings $\lambda_\phi$ and $\lambda_\phi^\prime$ acting as new spurion sources of baryon number violation.
}.

\begin{table}[tb]
  \begin{center}
 \begin{tabular}{c|c}
 Spurion & Leading term  \\
 \hline\hline
   $[\mathcal{S}_{N^*N^{\dagger}}]_S$ & $ \eL  $  \\
  $[\mathcal{S}_{N^*N^{\dagger}}]_A$ & $  Y_\nu^T Y_\nu^* \eL - \eL Y_\nu^\dag Y_\nu $  \\
  $[\mathcal{S}_{L^*L^{\dagger}}]_S  $  & $Y_{\nu}^* \epsilon_L Y_{\nu}^{\dagger}$
 \\
  $S_{X}$ & $Y_X$ \\
   $S_{XX^{\dagger}}$  & $\mathbb{1} $  \\
\end{tabular}
\end{center}
\caption{\label{tab:spurions_resume} Leading scaling of the  spurions analyzed in Sec.~\ref{sec:hierarchies_5} and Sec.~\ref{sec:hierarchies_6}, where $X=e,u,d,q,L,N$. All terms have generic $\mathcal{O}(1)$ factors that we do not write explicitly. }
 \end{table}

\section{Phenomenological implications}
\label{sec:pheno}

We have seen that MFV ans\"atz implies very different sizes for the coefficients of the effective operators reported in Tab.~\ref{tab:D6_operators}, and this has important consequences for present and future collider searches of RH neutrinos in the $[1\div 100]\,$GeV range, as well as on their interpretation in terms of a given model structure. The main result of the analysis in the previous section is that operators with two RH neutrinos that preserve lepton number,  that is ${\cal O}^6_{NX}$ with $X=e,u,d,q,L,H$, are the only ones that involve interactions with the SM particles and that can have coefficients of ${\mathcal O}(1)$ under the MFV hypothesis~\footnote{This feature was also previously pointed out in Ref.~\cite{Alonso:2011jd}.}. These are therefore the interactions that could compete with the active-sterile mixing  effects
 to enhance the production of RH neutrinos at colliders, that would necessarily  then be produced in pairs.  There is a different operator that contains two RH neutrinos, ${\cal O}^5_{NH}$. This breaks lepton number and has then an $\epsilon_L$ but no suppression in the Yukawa couplings.
 On the other hand, all operators that contain a single RH neutrino,
 and therefore could contribute to their decay, are suppressed by at least one power of $Y_\nu$, {\emph{i.e.}} they have the same parametric dependence of the active-sterile mixing $\theta_{\nu N}$. In this Section we will comment on how the MFV hypothesis influences searches in present and future colliders. A detailed analysis of their reach is beyond the scope of this paper and is left for future work~\cite{inprogress}. In Tab.~\ref{tab:spurions_resume} we summed up the scaling of the relevant spurions, which can be used to easily estimate the suppression under the MFV ans\"atz of any phenomenological search of interest.

\subsection{Exotic Higgs decay}
The dimension five ${\cal O}^5_{NH}$ operator gives rise to the exotic decay of the SM Higgs into a pair of right handed neutrinos. In Ref.~\cite{Caputo:2017pit} the authors investigated the reach of the Large Hadron Collider (LHC) to this interaction, ignoring the effect that dimension six operators could have on their decay lengths. The decay was assumed to be mediated by mixing and resulted in displaced  decays, which was an essential feature of the search strategy. Moreover, as long as the decay length in the laboratory frame $L \sim \theta_{\nu N}^2 M_N^6$ is in the ballpark range for displaced vertices searches at LHC, the bound was found to be essentially independent of the RH neutrino mass (clearly in the kinematic region wherethe Higgs decay channel is open).  In the most favorable scenario a  bound of $\Lambda \gtrsim 160$ TeV was estimated for   $300$ fb$^{-1}$ of integrated luminosity at LHC 13TeV.

Under the MFV hypothesis, the Wilson coefficient of this operator has an extra $\epsilon_L$ suppression. This weakens the limit on $\Lambda$, and introduces a stronger dependence on the RH neutrino mass $M_N$. We obtain 
\be\label{eq:Lambda_MFV_ONH5}
\Lambda \gtrsim 4~\mathrm{TeV} \sqrt{\frac{\bar{M}}{100~\mathrm{GeV}}}\ .
\ee

It is crucial that the production channel through the decay of the SM Higgs boson is not suppressed by any Yukawa insertion, as already foreseen in \cite{Graesser:2007yj}.  Any such suppression  would reduce the efficiency of the production mechanism for $N_R$, making it similar to that via mixing and beyond reach of LHC. 
We stress that even in the limit of degenerate RH neutrino masses the $\mathcal{O}_{NH}^5$ operator still violates lepton number. The MFV assumption  then requires its Wilson coefficient to have the same $\eL$ suppression considered above. This implies that also in this case the bounds on the scale $\Lambda$ are reduced to Eq.~\eqref{eq:Lambda_MFV_ONH5}.

On the other hand, the unsuppressed operators of dimension six,  ${\cal O}^6_{NX}$, could potentially provide a more efficient production mechanism,  as long as $\epsilon_L < {v\over \Lambda}$, as we discuss in the following.

\subsection{Pair production of RH neutrinos at future lepton facilities}

We first consider the future International Linear Collider (ILC) operating at a center of mass energy of $\sqrt{s} = 500\;$GeV. In Ref.~\cite{Antusch:2016vyf} the authors estimated the reach on the combination $|\theta_{e}|^2=\sum_{\alpha=1,2}|\theta_{e,\alpha}|^2$ to be or order $4 \times 10^{-9}$ for $M_N\simeq 50\,$GeV~\footnote{At energies well above the $Z$ pole the dominant contribution to the cross-section arises from the exchange of a t-channel $W$, hence the dependence of the results on $\theta_{\nu_e}$ only.}. This limit is obtained by assuming singly produced RH neutrinos through an s-channel $Z$ or t-channel $W$ and with a total integrated luminosity of $5\,{\rm ab}^{-1}$.
 For this value of the mixing angle the $e^+ e^-\to \nu N$ cross-section is $\sigma\simeq 8\times 10^{-4}$ fb~\cite{Antusch:2016vyf}. On the other hand the dimension-six operator ${\cal O}_{NL}$  gives a cross-section~\cite{Peressutti:2011kx}
\be
\sigma_{{\cal O}_{NL}}\simeq |{\cal S}_{NN^\dagger} {\cal S}_{LL^\dagger}|^2  \frac{s \beta}{64\pi^2 \Lambda^4}\left(1+\frac{\beta^2}{3}\right)\ , 
\ee
where $\beta=\sqrt{1-4M_N^2/s}$. If the RH neutrinos are long-lived~\footnote{For $|\theta_{e N}|^2\sim 4\times 10^{-9}$ and $M_N\sim30\,$GeV we get a decay length via mixing 
 of $\simeq 5\,$cm in the laboratory frame.}, this operator gives rise to a signature with a pair of displaced vertices, probably easy to be identified in the clean environment of a leptonic machine. By making the simplified, and perhaps conservative, assumption that the experimental sensitivity on the $e^+ e^- \to N N$ process is the same as the one for the $e^+ e^- \to \nu N$ process, {\emph{i.e.}} that we can exclude a cross-section of $\sigma_{{\cal O}_{LN}} \sim 8\times 10^{-4}\,$fb, we estimate that the ILC could test a scale up to $\Lambda\sim 22\,$TeV, thus surpassing the reach that one could obtain at the LHC from exotics Higgs decay via the $d=5$ operator.

\subsection{Searches at FCC-eh and FCC-hh}

It is interesting to note that $d=6$ operators built out with quarks bilinear could potentially give  observable effects at future electron-proton (FCC-eh) and proton-proton (FCC-hh) facilities~\cite{Abada:2019lih}, see {\emph{e.g.}}~\cite{delAguila:2008ir,Alcaide:2019pnf}. For what concerns FCC-eh, operators as ${\cal{O}}^6_{NLqu}$ could be tested in processes as $p e \to N q$, where $q$ represent any left- or right-handed quark. As pointed out in Sec.~\ref{sec:additional_spurions}, all these operators suffer from a double Yukawa insertion, one related to the neutrino sector and one to the quark sector, ending up being highly suppressed. We thus expect that they will  hardly be testable at this facility.
On the other hand FCC-hh could improve significantly the bounds to the unsuppressed operator ${\cal{O}}^6_{Nq}$ through, {\emph{e.g.}}, monojet processes $pp\to j N_R N_R$. Such a process was considered in~\cite{Alcaide:2019pnf} for the case of the LHC. In \cite{Alcaide:2019pnf}, a search of one lepton and missing transverse energy $u\bar{d} \rightarrow l_i^+ N_R$ was proposed to constrain the operator ${\cal{O}}^6_{NLqu}$, which however is Yukawa suppressed in the MFV hypothesis and therefore not competitive.

\begin{figure}[tb]
\begin{center}
\includegraphics[width=0.49\textwidth]{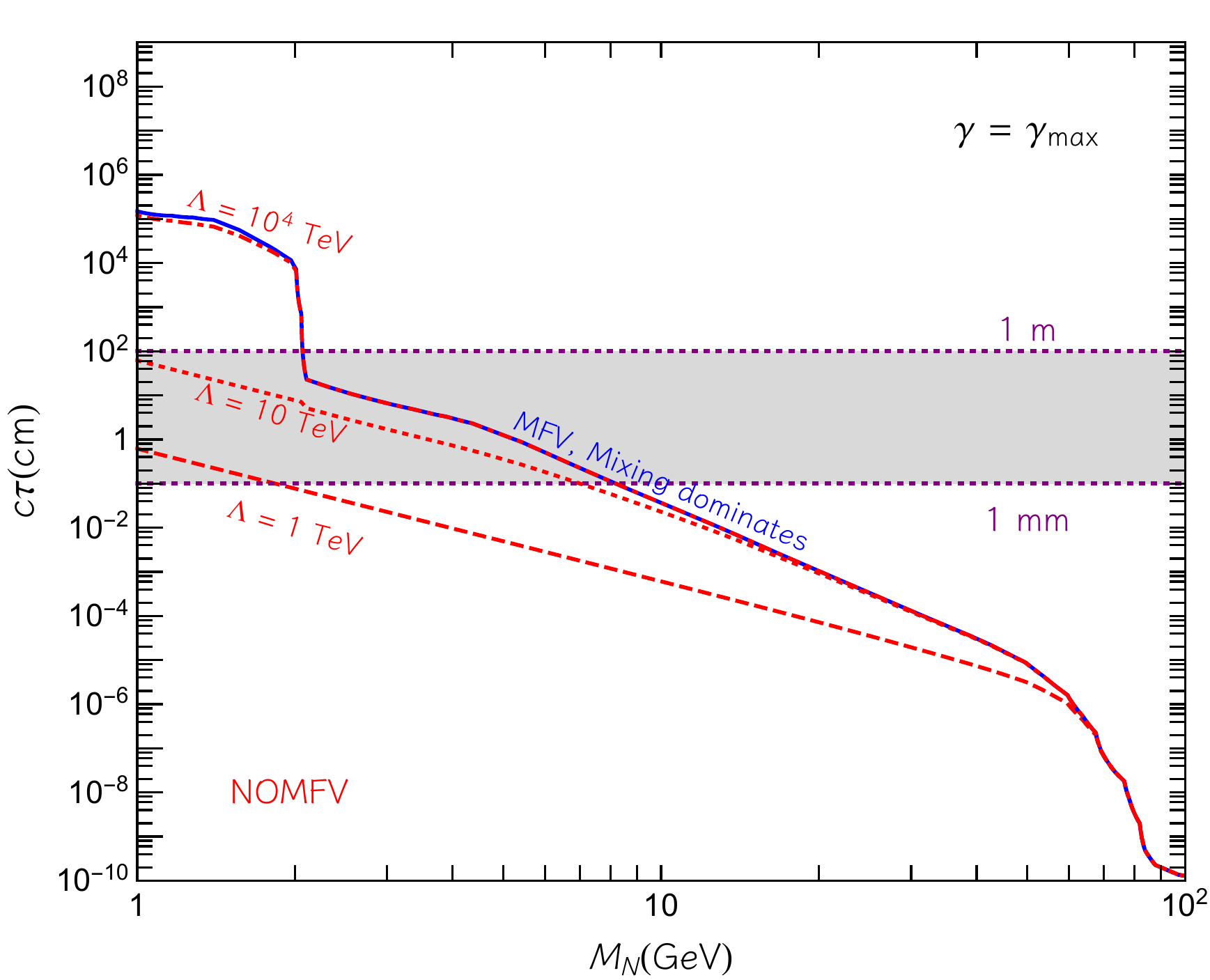}\hfill
\includegraphics[width=0.49\textwidth]{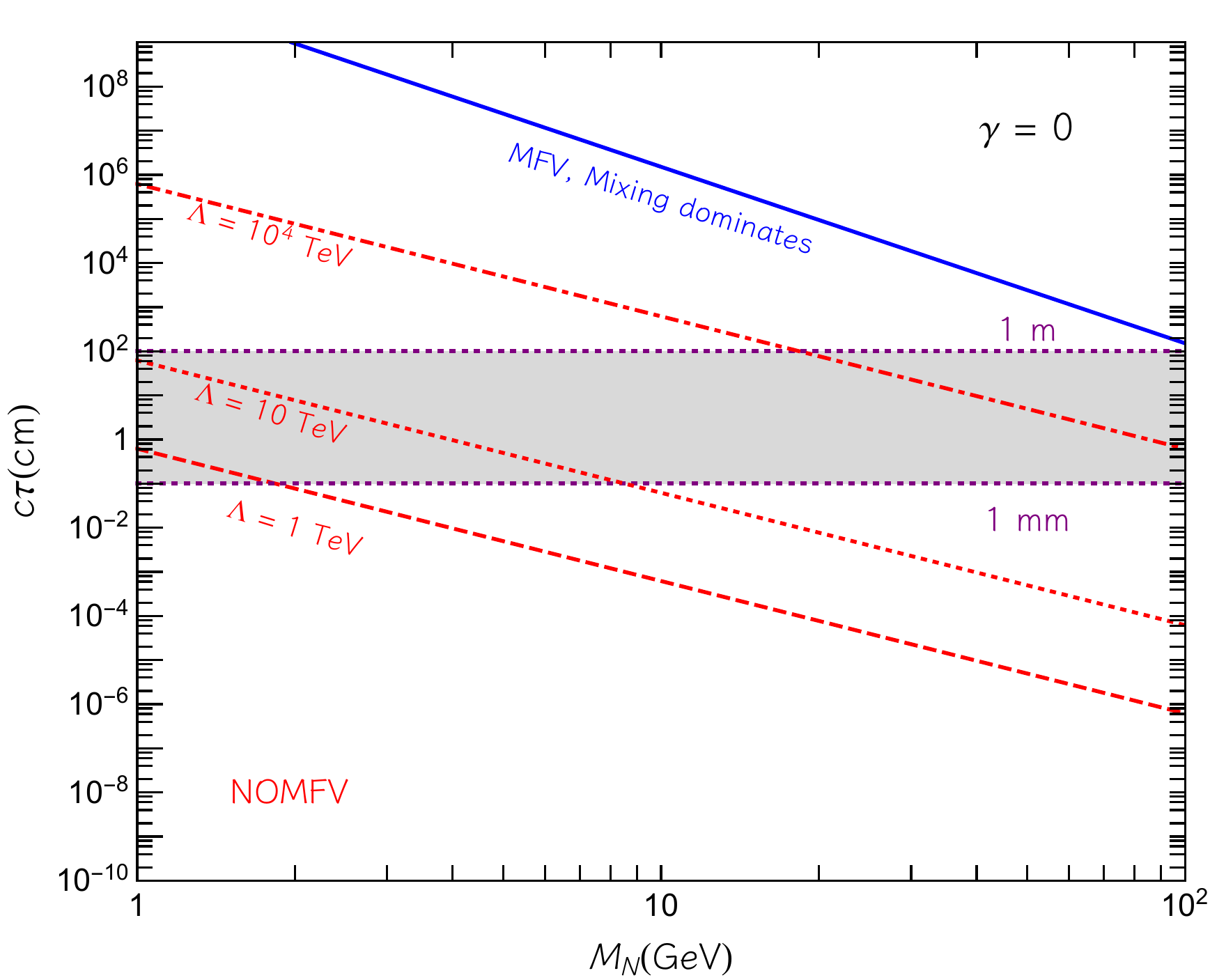}
\caption{\small Proper decay length $c\tau$ of the sterile neutrinos as a function of the mass scale $M_{N_{1,2}}$ for different choices of parameters: $\gamma = \gamma_{max}$ and $\alpha =0$ (left) and $\gamma = 0$ and $\alpha = \pi/4$ (right). The red lines show the results without assuming MFV for three different values of the new physics scale: $\Lambda = 1$ TeV (dashed line), $\Lambda = 10$ TeV (dotted line) and $\Lambda = 10^4$ TeV (dot-dashed line). The blue lines are instead  drawn assuming MFV, with the decay rate driven by the mixing. In the gray region the decay will produce a detectable displaced vertex at the LHC~\cite{Aad:2012zx}.}
\label{fig:decay}
\end{center}
\end{figure}

\subsection{Prompt and displaced decays}

A crucial consequence of the MFV ans\"atz regards the lifetime of the RH neutrino states. In the absence of higher dimensional operators, they decay through the mixing with the active neutrinos. The partial rate for $N_2$ to decay in the first generation of SM leptons can be approximated as~\cite{Atre:2009rg}
\begin{equation} \label{mixingrate}
	\Gamma_{\rm{\theta}} \simeq 10^{-2}~\mathrm{GeV}\, \Big(\frac{M_{N_2}}{100\,{\rm GeV}}\Big)^5 |\theta_{e,2}|^2 \ .
\end{equation}
As we saw in Fig.~\ref{fig:active_sterile_bounds}, the active-sterile mixing angle depends on both the RH neutrino masses and $\gamma$. For small $\gamma$ one obtains a proper decay length $c\tau > 0.1\,$cm for all values $M_{N_{1,2}} = [1 \div 100]$ GeV. In particular, for 
$M_{N_{2}}\lesssim 90\;$GeV, $c\tau > 1$ m, and most  RH neutrinos decay outside the detector, while for larger masses  $0.1\,\mathrm{cm} <c\tau<1\,\mathrm{m}$, and the decay is mostly displaced. As $\gamma$ increases, the interval in which the decay is likely to be displaced grows towards smaller RH neutrino masses, until we reach $\gamma \simeq 4$. For this value of $\gamma$ a window at large masses in which the decay is prompt opens up. For $\gamma \gtrsim 8$ (\textit{i.e.} for $M_{N_{1,2}} \gtrsim 10~\mathrm{GeV}$), all RH neutrinos in the target mass range decay promptly.

Higher dimensional operators that induce new decay modes for the RH neutrinos can drastically modify this behavior.  At $d=5$, the ${\cal O}^5_{NB}$ operator gives rise, if kinematically allowed, to the additional decay $N_2 \to N_1 \gamma$ with an estimated rate~\footnote{For simplicity, we neglect phase space suppression in this decay mode.}~
\be\label{eq:decayd5}
\Gamma_{{\cal O}^5_{NB}} \sim \frac{1}{4\pi}\frac{1}{(16\pi^2)^2}\frac{M^3_{N_2}}{\Lambda^2} \left| [{\mathcal S}_{N^* N^\dagger}]_A \right|^2\ .
\ee
where one should consider only the relevant entry of the spurion matrix and where we have also included the expected loop suppression factor as indicated in Tab.~\ref{tab:dim5}. The coefficient $|[S_{N^*N^\dagger}]_A|^2$ is generally of  ${\mathcal O}(1)$ under general assumptions, while the imposition of MFV implies the strong suppression $|[S_{N^*N^\dagger}]_A|^2 \propto {\mathcal O}(Y_\nu^2 \epsilon_L)^2$, as shown in Eq.~\eqref{eq:spurion_ONB5}. At $d=6$, the operator ${\cal O}^6_{L N B}$ (which is also loop-suppressed, see Tab.~\ref{tab:D6_operators}) allows for the decay $N_2 \to \nu_e \gamma$ with a rate that we estimate to be
\be\label{eq:decayd6}
\Gamma_{{\cal O}^6_{LNB}} \sim  \frac{1}{4\pi}\frac{1}{(16\pi^2)^2} \frac{v^2}{\Lambda^4}M_{N_2}^3 |{\mathcal S}_\nu|^2 .
\ee
Again, the entries of the spurions are generally ${\mathcal O}(1)$ while MFV implies $|{\mathcal S}_\nu|^2 \propto \mathcal O(Y_{\nu}^2)$.

In general, the decay rates induced by higher dimensional operators can easily dominate over the decay rate induced by mixing \cite{Butterworth:2019iff}, provided that the Wilson coefficients are ${\mathcal O}(1)$ and $\Lambda/v$ is not too large. The MFV hypothesis implies an additional suppression in the Wilson coefficients that results in the hierarchy $\Gamma_{{\cal O}^5_{NB}},\Gamma_{{\cal O}^6_{LNB}} \ll \Gamma_{\rm{\theta}}$, so that under MFV the dominant decay channel is via active-sterile mixing.  This is illustrated in  Fig.~\ref{fig:decay}, where we show $c\tau$ as a function of the RH neutrino mass for two extreme scenarios with widely different values of the active-sterile mixing, $\theta_{\nu N}$. On the left panel we assume that the mixing is as large as possible, \textit{i.e.} $\gamma=\gamma_{\rm max}$ compatible with present constraints (note that the upper limit discussed before depends on the RH neutrino mass). On the right panel we take $\gamma=0$. We compare the result assuming MFV, solid blue line, where the dominant decay arises via active-sterile mixing, with the ones with general ${\mathcal O}(1)$ Wilson coefficients
for different values of $\Lambda=1, 10$ and $ 10^4$ TeV. The horizontal band shows the values of $c\tau$ corresponding to displaced decays observable at LHC. 

As mentioned above, if the MFV ans\"atz is imposed the mixing always dominates over the higher dimensional operators and drives the decay. In this case the usual sterile neutrino searches~\cite{Caputo:2016ojx, Caputo:2017pit, Antusch:2015mia, Antusch:2016vyf} are not affected by the presence of higher dimensional operators, and the $c\tau$ does not depend on $\Lambda$. On the other hand, if MFV is not imposed, and $\Lambda$ is not too large, the higher dimensional operators dominate the decay in a large region of the parameter space, making $c\tau$ depend strongly on $\Lambda$.  

For the largest possible values of the mixings, $\gamma = \gamma_{max}$, left panel, we see that the range of masses for which displaced decays are expected  is between $[2 \div 10]$ GeV if MFV is assumed, while for larger masses prompt decays will occur. This region shifts to lower masses as $\Lambda$ decreases in the absence of MFV.   
For smaller mixings $\gamma\sim0$, right panel, decay via mixing always leads to  average decay lengths  much longer than the LHC detector sizes. However the situation changes dramatically with the presence of higher dimensional operators if no MFV is assumed: even for values of $\Lambda$ as large as $\Lambda \simeq 10^4$ TeV, the average decay length could correspond to displaced decays for the largest mass range and even prompt decay for smaller $\Lambda$. The effects of the higher dimensional operators only become negligible for scales of order $\Lambda \gtrsim 10^6$ TeV. 

Summing up, we have illustrated the two effects that modify  the pattern of the RH neutrino decays when more than one RH neutrino is added to the SM in the mass range $[1\div10^2]$ GeV, and higher dimensional operators are also considered. The first one is  the active-sterile neutrino mixing. While for small mixing,  $\gamma \simeq 0$, the decay length is always outside the detector,  for the largest values of $\gamma$ the decay can be displaced or even prompt. The situation is further modified when higher dimensional operators are considered. If we do not assume any symmetry principle, the additional channels opened up by $d=5$ and $d=6$ operators drive the decay length to smaller values. If MFV is imposed, the effect of the higher dimensional operators is negligible, and the decay is always dominated by the active-sterile mixing. Collider searches of displaced decays of RH neutrinos can thus be very useful to identify an underlying flavor structure of the theory.

\subsection{Astrophysics}

We now briefly comment on the implications of the MFV ans\"atz for astrophysical studies, relevant for slightly lower neutrino masses than those considered in the previous Sections~\cite{Raffelt:1987yt,Raffelt:1992pi,Raffelt:1996wa,Raffelt:1999gv,Haft:1993jt,Castellani:1993hs, Heger:2008er}.  The basic idea is that RH neutrinos in this mass range can modify stellar evolution, in particular non-degenerate stars and supernovae,  for neutrino masses $M_{N_{1,2}} \lesssim 10 \,$MeV.  

As first discussed in~\cite{Aparici:2009fh}, in the mass region $M_{\nu} \ll \,$ 10 keV, the $d=5$ dipole moment operator ${\cal O}_{NB}^5$ will produce a dominant decay $\gamma \rightarrow N_R N_R$ of a plasmon into two sterile neutrinos, resulting in the limit $\Lambda \gtrsim 4 \times 10^{6} $ TeV.  The same reasoning can be applied to supernovae bounds. The relevant mass range in this case is $10$ keV$ \, \lesssim M_{N_{1,2}} \lesssim \,$30 MeV, for which a new cooling process $\gamma + \nu \rightarrow N_R$ can occur, implying the lower limit \cite{Aparici:2009fh}
\begin{equation}
	\Lambda \gtrsim 4\times  10^{6}\sqrt{\frac{m_{\nu}}{M_{N_{1,2}}}}~ \mathrm{TeV}\ .
\end{equation}

These results assume ${\mathcal O}(1)$ Wilson couplings. Instead, if the MFV hypothesis is assumed,  the relevant operators are suppressed by the light neutrino mass and, as a result,  no meaningful constrain on the scale $\Lambda$ can be derived from these astrophysical observables.
\section{Conclusions} 
\label{sec:conclusions}

The evidence for non zero neutrino masses and mixings requires extending the SM with additional degrees of freedom. One of the simplest possibilities is to add to the SM particle content two or more RH neutrinos. Active neutrino masses compatible with experimental measurements are generated  by an interplay of the Yukawa coupling between the active and sterile neutrinos and the Majorana mass term for the new RH states via the see-saw relation. Motivated by considerations related to naturalness and the observation of a large baryon asymmetry in the Universe, we focused on RH neutrino masses between $[1 \div 100]\,$GeV, {\emph{ i.e.}} in a mass range relevant for present and future collider searches. In this mass range and in the absence of other new physics,  the RH neutrinos can be produced via mixing with the active neutrino states in charged and neutral current processes or Higgs decays. Also the decay of these particles is in this case driven by mixing via charged currents. The presence of additional new physics states at a scale $\Lambda \gg v, M_{N_{1,2}}$ can modify the phenomenology of the RH neutrinos, which therefore become a new portal, {\it the see-saw portal}. Such modifications can be parametrized at low energies as an effective field theory with higher dimensional operators ${\cal O}^{4+d}/ \Lambda^d$ with $d>4$, that include both the SM fields and two RH singlets. This effective theory has been subject of various studies before \cite{Graesser:2007pc,Graesser:2007yj,delAguila:2008ir,Aparici:2009fh,Liao:2016qyd}. 

In this work we have considered the implications of the MFV principle \cite{Cirigliano:2005ck,Davidson:2006bd,Branco:2006hz,Gavela:2009cd,Alonso:2011jd,Dinh:2017smk} in this theory. We have presented the dependence of the Wilson coefficients of the $d=5$ and $d=6$ operators involving RH neutrino fields on the flavor spurions parametrizing lepton flavor and lepton number breaking effects in the renormalizable Lagrangian, highlighting which ones are suppressed by the tiny active neutrino masses and which are not. Particular attention has been devoted to the most accessible parameter space that corresponds to large active-sterile mixing.  We have then discussed the most important phenomenological consequences  relevant for present and future collider experiments, particularly the aspects related to production rate and decay properties of the RH neutrinos, since the most sensitive searches are based on displaced decay patterns. In particular we have found that the imposition of the MFV ans\"atz can strongly modify previous estimates of the decay length of the RH neutrinos induced by $d=5$ and $d=6$ operators. In particular, our main result is that the imposition of the MFV hypothesis implies that the decay of the RH neutrinos is always dominated by mixing. On the other hand, we have found that pair production can have strongly enhanced production rates at colliders with respect to the single production mediated by  mixing, even if MFV is assumed, via $d=6$ operators of the form $\bar{N}_R \gamma^\mu N_R \bar{X} \gamma_\mu X$, with $X=Q, L, u,d, e$. Sensitivity of future colliders, such as FCC-ee, -eh and -hh, to $\Lambda$ will significanlty improve present LHC bounds. Finally, we have also shown the consequences of MFV in astrophysical searches, and found that they become non-competitive under this hypothesis.

\acknowledgments

DB thanks the Galileo Galilei Institute for theoretical physics for hospitality while part of this work was carried out. EB was supported by Funda\c{c}\~ao de Amparo \`a Pesquisa (FAPESP), under contract 2015/25884-4. AC acknowledges support from the Generalitat Valenciana (GVA) through the GenT program (CIDEGENT/2018/019). Furthermore AC and PH acknowledge support from the GVA project PROMETEO/2019/083,  as well as the national grant FPA2017-85985-P, and  the European projects H2020-MSCA-ITN-2015//674896-ELUSIVES and 690575-InvisiblesPlus-H2020-MSCA- RISE-2015.

\appendix

\bibliographystyle{h-physrev}
\bibliography{MFV_RHnus}

\end{document}